\documentclass[preprint,12pt,3p]{elsarticle}
\usepackage{color}
\definecolor{light-gray}{gray}{0.80}
\usepackage[]{listings}
\usepackage{xcolor}
\usepackage{amsfonts}
\usepackage{amsmath}
\usepackage{amssymb}
\usepackage{amsthm}
\usepackage{pgfgantt}
\usepackage{tikz}
\usepackage{caption}
\usepackage{subcaption}
\usepackage[colorlinks,citecolor=red,urlcolor=blue,bookmarks=false,hypertexnames=true]{hyperref} 
\usepackage[linesnumbered,ruled,vlined]{algorithm2e}
\usepackage{algorithmic}
\usepackage{multirow}
\usepackage{graphicx}
\usepackage{enumerate}
\usepackage{multirow}
\usepackage{tikz}
\tikzstyle{startstop} = [rectangle, rounded corners, minimum width=3cm, minimum height=1cm, align=center, text width=4cm, draw=black, fill=red!30 ]
\tikzstyle{io} = [trapezium, trapezium left angle=70, trapezium right angle=110, minimum width=3cm, minimum height=1cm, align=center, text width=4cm, draw=black, fill=blue!30]
\tikzstyle{process} = [rectangle, minimum width=3cm, minimum height=1cm, align=center, text width=4cm, draw=black, fill=orange!30]
\tikzstyle{process1} = [rectangle, minimum width=5cm, minimum height=1cm, align=center, text width=5cm, draw=black, fill=orange!30]
\tikzstyle{decision} = [diamond, minimum width=1cm, minimum height=1cm, align=center, text width=4cm, draw=black, fill=green!30]
\tikzstyle{arrow} = [thick,->,>=stealth]

\graphicspath{{Figures/}}

\begin{document}

\begin{frontmatter}


\title{An open-source implementation of a phase-ﬁeld model for brittle fracture using Gridap in Julia}



\author[]{Mohammad Masiur Rahaman\corref{cor1}}
\ead{masiurr@iitbbs.ac.in}
\address[]{School of Infrastructure, Indian Institute of Technology Bhubaneswar, 752050, India}
\cortext[cor1]{Corresponding author}

\begin{abstract}
This article proposes an open-source implementation of a phase-field model for brittle fracture using a recently developed finite element toolbox, Gridap in Julia. The present work exploits the advantages of both the phase-field model and Gridap toolbox for simulating fracture in brittle materials.  On one hand, the use of the phase-field model, which is a continuum approach and uses a diffuse representation of sharp cracks, enables the proposed implementation to overcome such well-known drawbacks of the discrete approach for predicting complex crack paths as the need for re-meshing, enrichment of finite element shape functions and an explicit tracking of the crack surfaces. On the other hand, the use of Gridap makes the proposed implementation very compact and user-friendly that requires low memory usage and provides a high degree of flexibility to the users in defining weak forms of partial differential equations. A test on a notched beam under symmetric three-point bending and a set of tests on a notched beam with three holes under asymmetric three-point bending is considered to demonstrate how the proposed Gridap based phase-field Julia code can be used to simulate fracture in brittle materials.
\end{abstract}


\end{frontmatter}


\section{Introduction}
\label{Sec:Introduction}
To design structures with high reliability, failure analysis of structures is of great importance in engineering applications. As fracture due to crack initiation and propagation is one of the most often encountered failure modes in engineering materials and structures, modeling of fracture in solids has always been one of the most intriguing topics of research interests. Numerical modeling of fracture in solids has mainly been done either by using a discrete or a continuum approach. In the discrete approach, cracks in the material body are modeled as the discontinuity of the displacements in the domain whereas in the continuum approach a diffused approximation of cracks is used to model fracture as a continuum damage process for which displacements are continuous but the material stiffness gradually degrades. Linear elastic fracture mechanics (LEFM) \cite{griffitli1920phenomena,irwin1956onset,williams2001introduction,luo2018linear} and cohesive zone model (CZM) \cite{barenblatt1962mathematical,dugdale1960yielding} are the notable theories in the category of the discrete approach for fracture modelling. Although LEFM and CZM are very popular, their implementation requires an explicit tracking of the discontinuity in the displacement field that poses difficulty in modeling an arbitrary complex crack path.

Knowing the well-known drawbacks of the discrete approach for modeling complicated crack paths, researchers generally refer to the continuum approach that provides the crack paths as part of the solutions of the governing partial differential equations. One of the most popular theories in the category of continuum approach is the phase-field model (PFM) \cite{ambati2015review}. There are of course several phase-field approaches to brittle fracture that have been independently developed in the mechanics community \cite{francfort1998revisiting, bourdin2000numerical,bourdin2008variational,kuhn2008phase,amor2009regularized,kuhn2010continuum,miehe2010phase,borden2014higher,dhas2018phase} and in the physics community \cite{aranson2000continuum,karma2001phase,hakim2009laws,spatschek2011phase,eastgate2002fracture,henry2004dynamic} as well. In this article, a phase-field model proposed by Dhas {\it{et al.}} \cite{dhas2018phase} is adopted as the model provides a thermodynamically consistent way of accommodating dissipative energy effects whenever needed. In all the phase-field models, a diffused approximation of sharp cracks is used by introducing a length scale parameter and an internal variable called phase-field. The accuracy of the diffused approximation depends on the value of the length scale parameter and may represent the original crack problem if the length scale parameter is chosen sufficiently small. Although this feature of phase-field models imposes a highly efficient implementation of the model while using the finite element method as very fine meshes are required for regularizing the sharp cracks with a small value of length scale parameter, the model has gained huge popularity in the research community as it can be incorporated in commercial finite element software such as Abaqus \cite{msekh2015abaqus,liu2016abaqus,molnar20172d, wu2020comprehensive,navidtehrani2021simple,navidtehrani2021unified}. However, to make the phase-field model available for a wider class of practitioners and researchers, there are also attempts towards open-source implementation of phase-field models by using finite element method \cite{natarajan2019fenics} and machine learning-based approaches \cite{samaniego2020energy,goswami2020transfer}. 

In this article, a new open-source implementation of a phase-field model is proposed using a recently developed finite element toolbox Gridap available in the programming language Julia \cite{bezanson2012julia,bezanson2017julia} that shares the advantages of both the static and dynamic languages. The programming language Julia is computationally efficient as the static languages such as Fortran, C++, etc., and also easy to use as the dynamic languages like Matlab, Python, Mathematica, etc. Gridap is an extensible finite element toolbox \cite{badia2020gridap,verdugo2019user} in Julia that can be used to solve a wide range of physical problems modeled mathematically using partial differential equations (PDEs). In contrast to other finite element libraries written in Julia such as FinEtools, JuAFEM, and JuliaFEM \cite{frondelius2017juliafem}, Gridap uses a novel software design (for example, high-level API
calls) that enables one to compute the value for a specific cell on the fly and never store the values for all cells in
the mesh simultaneously and thus essentially requires very low memory usage.  Moreover, Gridap provides a high degree of flexibility to the users as they can implement any PDEs-based mathematical model such as a phase-field model using a very compact syntax without explicitly writing any for-loop for assembly over elemental matrices. To develop a Gridap based open source program for phase-field modeling of brittle fracture, a thermodynamically consistent phase-field model is briefly described first in Section \ref{Sec:PhaseField}. Then, the derivation of the weak form corresponding to the governing PDEs of the phase-field model and the finite element implementation in Julia are provided in Section \ref{sec:JuliaImplementation}. Successful implementation of the phase-field model using Gridap is demonstrated in Section \ref{Sec:NumericalSimulation} through a  test on a notched beam under symmetric three-point bending and a set of tests on a notched beam with three holes under asymmetric three-point bending tests. Finally, the outcomes of the present work are summarized, and concluding remarks are accordingly made in Section \ref{sec:conclusion}.

\section{Phase-field model}
\label{Sec:PhaseField}
In this section, a brief description of a thermodynamically consistent phase-field approach \cite{dhas2018phase} to brittle fracture in elastic solids under small strains and isothermal conditions is provided.  

\subsection{Kinematics}
\begin{figure}[ht!]
  \subfloat[]{
	\begin{minipage}[c][0.5\width]{0.5\textwidth}
	   \centering
	 \includegraphics[width=\textwidth]{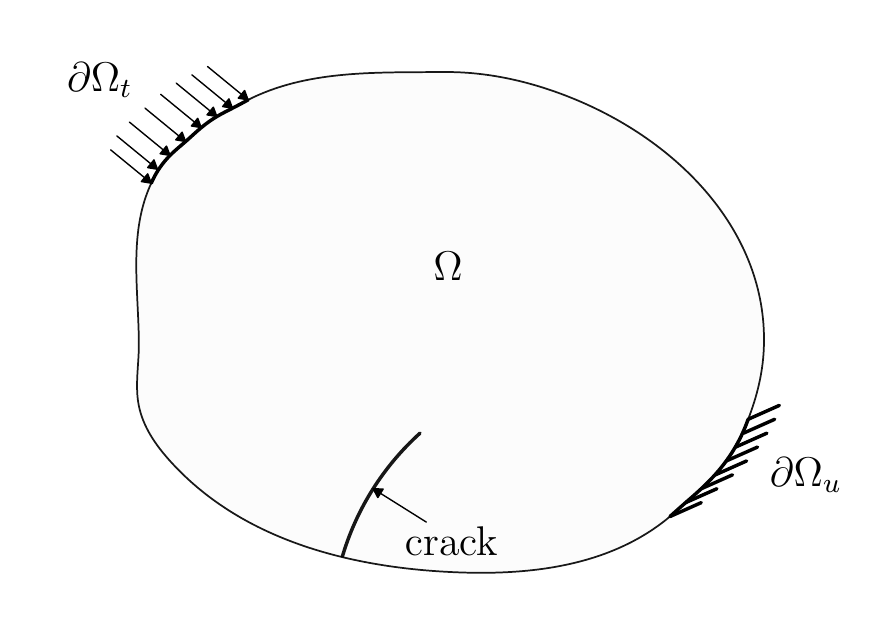}
	\end{minipage}}
  \subfloat[]{
	\begin{minipage}[c][0.5\width]{0.5\textwidth}
	  \centering
	 \includegraphics[width=\textwidth]{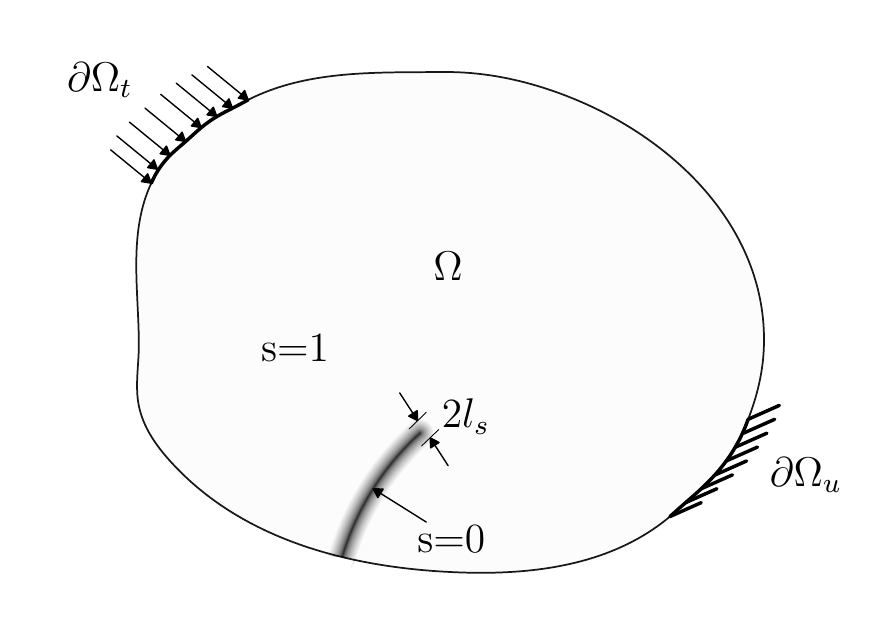}
	\end{minipage}}
\caption{A material body defined by an open set $\Omega$ is subjected to Dirichlet and applied traction boundary conditions on the boundaries $\partial \Omega_u$ and $\partial\Omega_t$ respectively. Sub-figures (a) and (b), respectively, show the material body with a sharp crack and a diffused representation of the crack via a regularization parameter $l_s$ and a scalar variable $s$. The variable $s$ varies from $0$ to $1$ with the value $0$ denoting fully damaged, $1$ undamaged and a value in between  partially damaged state.}
\label{Fig:PhaseField}
\end{figure}

Consider an open set $\Omega$ to be the reference configuration of a deformable body in the three dimensional Euclidean space $\mathbb{E}^3$. Let $\partial\Omega$ and $\bar{\Omega} = \Omega\cup\partial\Omega$ be the smooth boundary and the closure of $\Omega$, respectively. The displacement field may be defined as $\boldsymbol{u}(\cdot,t):\Omega\rightarrow\mathbb{R}^3$ at any instant of time $t\in\mathbb{R}^{+}$. Within a small deformation set-up, the strain tensor $\boldsymbol\epsilon$ is given by
\begin{equation}
\boldsymbol{\epsilon}(\boldsymbol u) = \frac{1}{2}(\nabla \boldsymbol u + \nabla \boldsymbol u^T),
\end{equation}
where $(\cdot)^T$ denotes transpose of a tensor and $\nabla$ the gradient operator. In the phase-field model, sharp cracks are approximated by using a diffused representation of the cracks via a length scale parameter $l_s$ and an internal variable $s$ called phase-field (see Fig. \ref{Fig:PhaseField}). Using the values of the phase-field variable, one can describe the  damaged, undamaged or partially damaged states of matter as follows: $s = 1$ for the undamaged state, $s = 0$ for the fully damaged state and $s\in(0,1)$ for a partially damaged state. Considering fully damaged states as the fracture, crack set may be defined as $\Omega_s = \{\boldsymbol{x} \in  \Omega|s(\boldsymbol{x}) = 0 \}$. Here, the damage process is considered to be irreversible that is if $\boldsymbol{x} \in \Omega_s$ at time $t_0$ then $\boldsymbol{x} \in \Omega_s$ for all $t \geq t_0$. Thus, the deformed and damaged states of the material body may be described by considering phase-field variable $s$ as an additional kinematic descriptor along with the displacement vector $\boldsymbol{u}$.

\subsection{Force Balances}
  To describe the deformation of a material body under external loading, one can derive the force balances from a virtual power principle by considering a macro- and a micro-force system  \cite{gurtin1996generalized}. While  stress tensor $\boldsymbol{\sigma}$, traction vector $\boldsymbol t(\boldsymbol n)$ and body force $\boldsymbol b$ define the macro-force system, the micro-force system includes scalar micro-traction  $\chi(\boldsymbol n)$, vector micro-stress $\boldsymbol{\xi}$ and scalar micro-stress $\pi$. It is important to note that the micro-system is also at the same continuum level of the macro-system and the coinage `micro-force' may be a misnomer. One can identify $\dot{\boldsymbol{u}}$ as the power conjugate of  $\boldsymbol t(\boldsymbol n)$ and $\boldsymbol b$, $\dot{s}$ as the power conjugate of $\chi(\boldsymbol n)$ and express the external power $P ^{\text{ext}}$ for any arbitrary part of the body $\mathcal{P}\subset\Omega$ as
  \begin{equation}
      P^{\text{ext}} = \int_{\partial\mathcal{P}}\biggl[\boldsymbol t \left(\boldsymbol n\right)\cdot \dot {\boldsymbol u} + \chi(\boldsymbol n) \cdot \dot s \biggr]\,dA+ \int_{\mathcal{P}} \boldsymbol{b} \cdot \dot{\boldsymbol{u}} \,dV,
       \label{eq:ExtPower}
  \end{equation}
 where $\dot{()}$ denotes the time derivative of a variable, $\boldsymbol{n}$ the unit normal vector to the boundary $\partial\mathcal{P}$, $dA$ and $dV$ are, respectively, measures on $\partial\mathcal{P}$ and $\mathcal{P}$. The internal power $P^{\text{int}}$ can be defined by the summation of power expenditure of $\boldsymbol{\sigma}$ over $\nabla \dot {\boldsymbol u}$, $\boldsymbol{\xi}$ over $\nabla \dot {s}$ and $\pi$ over $\dot{s}$, and can be given by
    \begin{equation}
         P^{\text{int}} = \int_{\mathcal{P}} \left(\boldsymbol{\sigma}:\nabla \dot {\boldsymbol u} + \boldsymbol{\xi}\cdot\nabla \dot {s} + \pi \dot{s}  \right) dV.
     \label{eq:IntPower}
    \end{equation}
Denoting the virtual counterparts of $\dot{\boldsymbol{u}}$ and $\dot{s}$ by $\tilde{\boldsymbol{u}}$ and $\tilde{s}$, respectively, one can define a set $\mathcal{V} =\left(\tilde{\boldsymbol{u}},\tilde{s}\right)$ called as the generalized virtual velocity vector. Then the external virtual power $\tilde{P}^{\text{ext}}\left(\mathcal{P};\mathcal{V}\right)$ and the internal virtual power $\tilde{P}^{\text{int}}\left(\mathcal{P};\mathcal{V}\right)$ can be expressed as
    \begin{equation}
    \tilde{P}^{\text{ext}}\left(\mathcal{P};\mathcal{V}\right) = \int_{\partial\mathcal{P}}\biggl[\boldsymbol t \left(\boldsymbol n\right)\cdot \tilde {\boldsymbol  u} + \chi(\boldsymbol n) \cdot  \tilde s \biggr]dA+ \int_{\mathcal{P}} \boldsymbol{b} \cdot  \tilde{\boldsymbol{ u}} \,dV,
    \label{eq:VirtExtPow}
    \end{equation}
    and
    \begin{equation}
     \tilde{P}^{\text{int}}\left(\mathcal{P};\mathcal{V}\right)= \int_{\mathcal{P}} \left(\boldsymbol{\sigma}:\nabla \tilde {\boldsymbol u} + \boldsymbol{\xi}\cdot\nabla \tilde {s} + \pi \tilde{s} \right) dV,
     \label{eq:VirtIntPow}
    \end{equation}
    respectively. Employing Eq. \eqref{eq:VirtExtPow},
    Eq. \eqref{eq:VirtIntPow} and invoking the virtual power principle i.e., $ \tilde{P}^{\text{ext}}\left(\mathcal{P};\mathcal{V}\right) =  \tilde{P}^{\text{int}}\left(\mathcal{P};\mathcal{V}\right)$, one can arrived at
    \begin{equation}
      \int_{\partial\mathcal{P}}\biggl[\boldsymbol t \left(\boldsymbol n\right)\cdot \tilde {\boldsymbol  u} + \chi(\boldsymbol n) \cdot \tilde s \biggr]dA+ \int_{\mathcal{P}} \boldsymbol{b} \cdot  \tilde{\boldsymbol{ u}} \,dV =  \int_{\mathcal{P}} \left(\boldsymbol{\sigma}:\nabla \tilde {\boldsymbol u} + \boldsymbol{\xi}\cdot\nabla \tilde {s} + \pi \tilde{s}  \right) \,dV.
      \label{eq:VirtWorkBal}
    \end{equation}
Appropriate choices of $\mathcal{V} =\left(\tilde{\boldsymbol{u}},\tilde{s}\right)$ in Eq. \eqref{eq:VirtWorkBal} may lead to the macro- and the micro-force balances as described below. 

\subsubsection{Macro-force balance}
    \label{subs:Macroscopic_Force_Balance}
Considering $\mathcal{V} =\left(\tilde{\boldsymbol{u}},0\right)$ i.e., by substituting $\tilde{s} = 0 $ in Eq. \eqref{eq:VirtWorkBal}, the macro-force balance equation may be obtained as
    \begin{equation}
         \int_{\partial\mathcal{P}}\left(\boldsymbol t \left(\boldsymbol n\right) -\boldsymbol{\sigma}\boldsymbol{n}\right).\tilde {\boldsymbol u}\, dA= -\int_{\mathcal{P}} \left(\nabla \cdot\boldsymbol{\sigma} + \boldsymbol{b}\right) \cdot \tilde {\boldsymbol{u}}\, dV,
         \label{eq:InvokeTildesZero}
    \end{equation}
which holds for all $\tilde{\boldsymbol{u}}$ and any arbitrary sub-domain $\mathcal{P}$. Applying the localization theorem on Eq. \eqref{eq:InvokeTildesZero}, one can get that 
    \begin{equation}
        \boldsymbol t \left(\boldsymbol n\right) = \boldsymbol{\sigma}\boldsymbol{n},
    \label{eq:MacroTractionCond}
    \end{equation}
which is the macro-traction condition and
    \begin{equation}
        \nabla \cdot\boldsymbol{\sigma} + \boldsymbol{b} = 0.
        \label{eq:MacroforceBal}
    \end{equation}
called the macro-force balance. Equations \eqref{eq:MacroTractionCond} and \eqref{eq:MacroforceBal} may be identified as the classical traction condition and the local linear momentum balance equation, respectively.

\subsubsection{Micro-force balance}
Considering $\mathcal{V} =\left(\mathbf{0},\tilde{s}\right)$ i.e., by substituting $\tilde{\boldsymbol {u}} = \mathbf{0}$ in Eq.\eqref{eq:VirtWorkBal}, the equation for micro-force balance can be derived as
      \begin{equation}
         \int_{\partial\mathcal{P}}\biggl[\chi(\boldsymbol n) - \xi \boldsymbol \cdot \boldsymbol{n}\biggr]\cdot \tilde{s} \,dA = -\int_{\mathcal{P}} \left(\nabla \cdot \boldsymbol{\xi} - \pi\right)\cdot \tilde{s}  \,dV.
         \label{eq:InvokeVirtTildeuZero}
      \end{equation}
 Since Eq. \eqref{eq:InvokeVirtTildeuZero} holds for all $\tilde{s}$ and any sub-domain $\mathcal{P}$, Eq. \eqref{eq:InvokeVirtTildeuZero} may be localized as 
      \begin{equation}
          \chi(\boldsymbol n) = \boldsymbol{\xi} \cdot \boldsymbol{n},
          \label{eq:MicroTractionCond}
      \end{equation}
      and 
      \begin{equation}
          \nabla \cdot \boldsymbol{\xi} - \pi = 0.
          \label{eq:MicroforceBal}
      \end{equation}
Equations \eqref{eq:MicroTractionCond} and \eqref{eq:MicroforceBal} are called the micro-traction condition and the micro-force balance, respectively.
      
\subsubsection{Thermodynamics and constitutive modeling}
In this section, constitutive relations for the macro- and micro- stresses are derived by imposing the first and second laws of thermodynamics. Considering an iso-thermal condition, one may state the second law of thermodynamics for any sub-domain $ \mathcal{P} \subset \Omega$ as the free energy inequality:
     \begin{equation}
         \frac{d}{dt} \int_\mathcal{P} \psi dV \le P^{\text{ext}},
         \label{eq:Second_law_of_thermodynamic}
     \end{equation}
where $\psi$ is the Helmholtz free-energy of the system. Using Eq. \eqref{eq:ExtPower}, Eq. \eqref{eq:IntPower} and the power balance, i.e. $P^\text{ext}$ = $P^\text{int} $, the inequality given by Eq. \eqref{eq:Second_law_of_thermodynamic} may be expressed as
\begin{equation}
    \begin{split}
         \int_\mathcal{P}\dot{\psi}dV \le \int_\mathcal{P}(\boldsymbol{\sigma}:\dot{\boldsymbol{\epsilon}} + \boldsymbol{\xi}\cdot\nabla \dot {s} + \pi \dot{s})\,dV,
    \end{split}
         \label{eq:Second_law_in_term_of_Int_Energy}
         \end{equation}
where the equality 
$\boldsymbol{\sigma}:\nabla\dot{\boldsymbol{u}} = \boldsymbol{\sigma}:\dot{\boldsymbol{\epsilon}}$ (which follows from the symmetry of $\boldsymbol{\sigma}$) is used. Since, the inequality given by Eq. \eqref{eq:Second_law_in_term_of_Int_Energy} holds for any arbitrary sub-domain $\mathcal{P}$, one can have
\begin{equation}
    \begin{split}
        \dot{\psi}-\left(\boldsymbol{\sigma}:\dot{\boldsymbol{\epsilon}} + \boldsymbol{\xi}\cdot\nabla \dot {s} + \pi \dot{s}\right)\le 0,
    \end{split}
    \label{eq:Energy_imbalance_in_strain_energy}
     \end{equation}
which must be satisfied whilst determining or postulating the constitutive relations for the thermodynamic fluxes $\boldsymbol{\sigma}$, $\boldsymbol{\xi}$ and $\pi$ in terms of the kinematic quantities $\boldsymbol{\epsilon}$, $\nabla s$ and $s$.

\subsection{Constitutive response functions}
\label{Constitutive response functions}
Let the free energy of the system $\psi$ be function of $\boldsymbol{\epsilon}$, $\nabla s$ , $s$ and may be written as
\begin{equation}
    \psi = \hat{\psi}( \boldsymbol{\epsilon},\nabla s , s).
    \label{eq:TotalFreeEnergy}
\end{equation}
Using the chain rule in Eq. \eqref{eq:TotalFreeEnergy}, one can get the rate of free energy as
\begin{equation}
    \dot{\psi} = \partial_{\boldsymbol{\epsilon}}{\psi}:\dot{\boldsymbol{\epsilon}}+  \partial_{\nabla s}{\psi}\cdot\nabla \dot{s} + \partial_{s}{\psi}\,\dot{s},
    \label{eq:Free_energy_derivatives}
\end{equation}
where $ \partial_{(.)} $ with a suffix represents the derivative of a function with respect to the argument in the suffix while keeping others
fixed. Considering that the scalar micro-stress $\pi$ has an energetic part $\pi^{\text{en}}$ and a dissipative part $\pi^{\text{dis}}$ i.e., $\pi = \pi^{\text{en}} + \pi^{\text{dis}} $
and substituting Eq. \eqref{eq:Free_energy_derivatives} in Eq. \eqref{eq:Energy_imbalance_in_strain_energy}, one can get that
\begin{equation}
    (\boldsymbol{\sigma}-\partial_{\boldsymbol{\epsilon}}\psi):\dot{\boldsymbol{\epsilon}}
   + (\boldsymbol{\xi}-\partial_{\nabla s}\psi)\cdot\nabla\dot{s}
   + (\pi^{\text{en}}-\partial_{s}\psi)\dot s + \pi^{\text{dis}}\dot s\ge 0.
 \label{eq:Energy_inequality_after_Heltmotz_energy}
\end{equation}
 Applying the Coleman-Noll procedure \cite{coleman1974thermodynamics} to Eq. \eqref{eq:Energy_inequality_after_Heltmotz_energy}, one can
  arrive at the constitutive relations for the thermodynamics fluxes as 
\begin{equation}
\boldsymbol{\sigma}=\partial_{\boldsymbol{\epsilon}}\psi,
\label{eq:Stress-constitutive-law}
\end{equation}
\begin{equation}
    \boldsymbol{\xi} = \partial_{\nabla s}{\psi},
    \label{eq:fracture-microstress}
\end{equation}
\begin{equation}
 \pi^{\text{en}} = \partial_{s}{\psi},
 \label{eq:Scalar-microstress}
\end{equation}
which leads to 
\begin{equation}
\pi^{\text{dis}}\dot {s}\ge 0.
\label{eq:piDis}
\end{equation}
Determination of the constitutive relation for $\pi^{\text{dis}}$ must be done in such a way that the inequality constraint given by Eq. \eqref{eq:piDis} always satisfy. From the irreversiblity condition on damage i.e. $\dot{s}\le 0$, it can be seen that a possible choice for  $\pi^{\text{dis}}$ could be $\pi^{\text{dis}} = \mathcal{G}\dot{s}$, where $\mathcal{G}$ is a constitutive function with $\mathcal{G}\le0$.
      
\subsection{Specialized constitutive relations}
To quantify the thermodynamic forces, one need to specialize the constitutive relations by postulating an explicit expression of the Helmholtz free energy $\psi$ in terms of $\boldsymbol{\epsilon}$, $\nabla s$ and $s$. Let the Helmholtz free energy $\psi(\boldsymbol{\epsilon},\nabla s,s)$ be a sum of elastic energy $\psi^{\text{elas}}(\boldsymbol{\epsilon},s)$ and fracture energy $\psi^{\text{frac}}(\nabla s,s)$ as
   \begin{equation}
       \psi(\boldsymbol{\epsilon},\nabla s,s) = \psi^{\text{elas}}(\boldsymbol{\epsilon},s) + \psi^{\text{frac}}(\nabla s,s). \label{eq:Total Energy}
   \end{equation}
  It is assumed that crack cannot propagate under pure compression and imposed by considering an additive decomposition of $\boldsymbol{\epsilon}$ into a volumetric part $\boldsymbol{\epsilon}_{\text{vol}}$ and a deviatoric part $\boldsymbol{\epsilon}_{\text{dev}}$ i.e.
   \begin{equation}
       \boldsymbol{\epsilon} = \boldsymbol{\epsilon}_{\text{vol}} + \boldsymbol{\epsilon}_{\text{dev}},
   \end{equation}
   where
   \begin{equation}
       \boldsymbol{\epsilon}_{\text{vol}} = \mathbb{P}_{\text{vol}}\,\boldsymbol{\epsilon};\,\,\,\,\, \boldsymbol{\epsilon}_{\text{dev}} = \mathbb{P}_{\text{dev}}\,\boldsymbol{\epsilon}.
       \label{eq:VolumetricAndDeviatoric}
   \end{equation}
  In Eq. \eqref{eq:VolumetricAndDeviatoric}, volumetric and deviatoric parts of a second order tensor are obtained by introducing fourth order projection tensors $\mathbb{P}_{\text{vol}}$ and $\mathbb{P}_{\text{dev}}$, respectively. Defining $p = \frac{1}{3}\boldsymbol{I}:\mathbb{P}_{\text{vol}}\mathbb{C}\boldsymbol{\epsilon}$ with $\mathbb{C}$ denoting the fourth order elasticity tensor, the elastic part of the free energy may be postulated as
   \begin{equation}
      \psi^{\text{elas}}\left(\boldsymbol{\epsilon}\right) = s^2\psi^{\text{elas}}_+\left(\boldsymbol{\epsilon}\right) + \psi^{\text{elas}}_-\left(\boldsymbol{\epsilon}\right),
      \label{eq:ElastEnerg}
   \end{equation}
where $\psi^{\text{elas}}_+\left(\boldsymbol{\epsilon}\right)$ is the elastic energy part due to a combination of pure tension and shear, 
 \begin{equation}
      \psi^{\text{elas}}_+\left(\boldsymbol{\epsilon}\right) = \frac{1}{2}\bigl(\langle p\rangle _+\boldsymbol{I}:\boldsymbol{\epsilon}_{\text{vol}} + \mathbb{P}_{\text{dev}}\mathbb{C}\boldsymbol{\epsilon}: \boldsymbol{\epsilon}_{\text{dev}}\bigr),
      \label{eq:PosPartElastEnerg}
  \end{equation}
and $\psi^{\text{elas}}_-\left(\boldsymbol{\epsilon}\right)$ is the elastic energy part due to pure compression
\begin{equation}
\psi^{\text{elas}}_-\left(\boldsymbol{\epsilon}\right) =\frac{1}{2}\langle p\rangle _-\boldsymbol{I}:\boldsymbol{\epsilon}_{\text{vol}}.
\label{eq:NegPartElastEnerg}
\end{equation}
In Eq. \eqref{eq:ElastEnerg}, a degradation function $s^2$ is introduced to account for the reduced stiffness of the material due to damage. Note that the degradation function $s^2$ is only associated with the so-called positive part of the elastic energy $\psi^{\text{elas}}_+\left(\boldsymbol{\epsilon}\right)$ to impose the condition that cracks cannot propagate under pure compression. In Eq. \eqref{eq:PosPartElastEnerg},  
$\langle p\rangle _+ = \frac{1}{2}\left(p +\lvert p\rvert \right)$ and in Eq. \eqref{eq:NegPartElastEnerg}  $\langle p\rangle _- = \frac{1}{2}\left(p-\lvert p\rvert \right) $.  The fracture energy $\psi^{\text{frac}}(\nabla s,s)$ may be postulated as
   \begin{equation}
       \psi^{\text{frac}}\left(\nabla s,s\right) = G_c\biggl(\frac{\left(1-s\right)^2}{2l_s} + \frac{l_s}{2}\nabla s \cdot  \nabla s\biggr),
      \label{eq:fracEnergy}
   \end{equation}
    where $G_c$ is the critical energy release rate and $l_s$ is the phase-field length scale parameter. Employing equations \eqref{eq:Stress-constitutive-law}, \eqref{eq:fracture-microstress}, \eqref{eq:Scalar-microstress}, \eqref{eq:PosPartElastEnerg}, \eqref{eq:NegPartElastEnerg} and  \eqref{eq:fracEnergy}, explicit expressions for the thermodynamic fluxes $\boldsymbol{\sigma}$, $\boldsymbol{\xi}$ and $\pi$ may be derived as
      \begin{equation}
          \boldsymbol{\sigma} = \partial_{\boldsymbol{\epsilon}}\psi = s^2\bigl({\langle p\rangle _+}\boldsymbol I + \mathbb{P}_{\text{dev}}\mathbb{C}\boldsymbol{\epsilon}\bigr) + \langle p\rangle _-\boldsymbol I,
          \label{eq:FinalExpSigma}
      \end{equation}
      \begin{equation}
          \boldsymbol{\xi} = \partial_{\nabla s}\psi = G_cl_s\nabla s,
      \end{equation}
      and
      \begin{equation}
          \pi^{\text{en}}  = \partial_s\psi  = 2s\psi^{\text{elas}}_+(\boldsymbol{\epsilon}^{\text{elas}})-\frac{G_c}{l_s}\left(1-s\right),
      \end{equation}
      respectively. Substituting the above constitutive relations in the macro- and micro-force balances, one may express the governing partial differential equations in terms of the kinematic descriptors $\boldsymbol{u}$ and $s$.

 \subsection{Boundary value problem and the strong form}
      \label{sec:BVPdef}
     Using the expressions derived in the previous section, one may write the strong form of the governing PDEs as a boundary value problem and derive the corresponding weak form of the governing PDEs for the finite element formulation of phase-field model. One can express the stress tensor $\boldsymbol{\sigma}$ in terms of kinematic variables as $\mathbb{C}_{mod}\, \boldsymbol{\epsilon}$ by defining $\mathbb{C}_{mod}$ as
\begin{equation}
 \mathbb{C}_\text{mod} = 
 \begin{cases}
    s^2 \mathbb{C},\,\,\,\,\,\text{for}\,\,p \ge 0  & \\
    s^2\left(\mathbb{P}_{\text{dev}}\mathbb{C}\right) + \mathbb{P}_{\text{vol}}\mathbb{C}, \,\,\,\,\text{for}\,\,p < 0.
  \end{cases}
  \label{eq:CtensorMod}
 \end{equation}
Using $\boldsymbol{\sigma} = \mathbb{C}_{mod}\, \boldsymbol{\epsilon}$, Eq. \eqref{eq:MacroforceBal} may be re-written as
      \begin{equation}
          \nabla \cdot\biggl(\mathbb{C}_{mod}\, \boldsymbol{\epsilon}\biggr) + \boldsymbol{b} = 0.
          \label{eq:ElastStrongForm}
      \end{equation}
Similarly, using the expressions of $\boldsymbol{\xi}$ and $\pi^{\text{en}}$ and assuming $\pi^{\text{dis}} = 0$, Eq. \eqref{eq:MicroforceBal} may be re-written as 
      \begin{equation}
          \nabla \cdot \biggl(G_cl_s\nabla s\biggr) - 2s\,\psi^{\text{elas}}_+(\boldsymbol{\epsilon})+\frac{G_c}{l_s}\left(1-s\right) = 0.
          \label{eq:RawFracStrongForm}
      \end{equation}
To account for the irreversiblity condition on damage, a history function $\mathcal{H}(\mathcal{E})$, where  $\mathcal{H}(f) = \text{max}_{\tau \in [0,t]}f(\tau)$ for any input argument $f$, of the so-called positive part of elastic energy $\mathcal{E} =\psi^{\text{elas}}_+(\boldsymbol{\epsilon})$ is employed \cite{miehe2010phase}. Using the history function $\mathcal{H}(\mathcal{E})$, Eq. \eqref{eq:RawFracStrongForm} may be expressed as
      \begin{equation}
          \nabla \cdot \biggl(G_cl_s\nabla s\biggr) - 2s\mathcal{H}(\mathcal{E})+\frac{G_c}{l_s}\left(1-s\right) = 0.
          \label{eq:FracStrongForm}
      \end{equation}
     The governing PDEs \eqref{eq:ElastStrongForm} and \eqref{eq:FracStrongForm} are coupled and subject to boundary conditions, such as  prescribed displacement $\boldsymbol{u} = \bar{\boldsymbol{u}}$ and applied traction $\boldsymbol{t}(\boldsymbol{n}) = \bar{\boldsymbol{t}}(\boldsymbol{n})$ on $\partial\Omega_u$ and  $\partial\Omega_t$, respectively. Equations \eqref{eq:ElastStrongForm} and \eqref{eq:FracStrongForm} together with the boundary conditions are called the strong form of the governing PDEs. In absence of body force i.e. $\boldsymbol{b}=\bf{0}$, the strong form for phase-field modeling of brittle fracture in an elastic solid defined by domain $\Omega$ subjected to displacement boundary condition $\boldsymbol{u} = \bar{\boldsymbol{u}}$ on the boundary $\partial\Omega_u$ can be given in a compact form as 
 \begin{subequations}
 \begin{align}
     \nabla \cdot \boldsymbol{\sigma} = {\bf{0}} \,\,\,\,\,\left(\text{where}\,\,\boldsymbol{\sigma} = \mathbb{C}_{mod}\, \boldsymbol{\epsilon}\right)\,\,\,\,\text{in}\,\,\,\,\Omega,\\
\boldsymbol{u} = \bar{\boldsymbol{u}} \,\,\,\,\text{on}\,\,\,\,\partial\Omega_u,\\
 \boldsymbol{\sigma}\boldsymbol{n} = {\bf{0}}\,\,\,\,\text{on}\,\,\,\,\partial\Omega\backslash\partial\Omega_u,
\end{align}
\label{eq:StrongFormDisplacement}
\end{subequations}
 corresponding to the macro-system and
 \begin{subequations}
 \begin{align}
    \nabla \cdot \biggl(G_cl\boldsymbol{A}\nabla s\biggr) - 2s\mathcal{H}(\mathcal{E})+\frac{G_c}{l}\left(1-s\right) = 0\,\,\,\,\text{in}\,\,\,\,\Omega,\\
    \nabla s \cdot\boldsymbol{n} = 0 \,\,\,\,\text{on}\,\,\,\,\partial\Omega\backslash\partial\Omega_u,
 \end{align}
 \label{eq:StrongFormPhaseField}
\end{subequations}
corresponding to the micro-system. In the present study, beams are made of isotropic materials for which components of the fourth order elasticity tensor may be expressed as $\mathbb{C}_{ijkl} = \lambda\delta_{ij}\delta_{kl} + \mu(\delta_{ik}\delta_{jl}+\delta_{il}\delta_{jk})$, where $\lambda$ and $\mu$ are the Lam\'e parameters and $\delta$ denotes the Kronecker delta. The Lam\'e parameters are related to Young's modulus $E$ and Poisson's ratio $\nu$ by $\lambda = E\nu/\left((1 + \nu)(1 - 2\nu)\right)$ and $\mu = E/\left(2(1 + \nu)\right)$.

\section{The weak form and an outline for finite element implementation in Julia}
  \label{sec:JuliaImplementation}
  
  In this section, first the weak form for the strong form given by Eq. \eqref{eq:StrongFormDisplacement} and Eq. \eqref{eq:StrongFormPhaseField} is derived and then an outline for the finite element formulation is provided through a numerical example. Let $\boldsymbol{v}$ and $\phi$ be the test functions corresponding to displacement $\boldsymbol{u}$ and phase-field $s$, respectively. The trial spaces with the given displacement boundary condition may be given by
      \begin{equation}
          H_u = \{\boldsymbol{u} \in H^1(\Omega) ; \; \boldsymbol{u} = \bar{\boldsymbol{u}} \; \text{on} \; \partial\Omega_u \},
         \label{eq:TrialSpaceU}
      \end{equation}
      \begin{equation}
          H_s = \{s \in H^1(\Omega) \},
      \end{equation}
where $\bar{\boldsymbol{u}}$ is a prescribed displacement on $\partial\Omega_u$. The test spaces may be defined as
    \begin{equation}
          V_u = \{\boldsymbol{v} \in H^1(\Omega); \; \boldsymbol{v} = 0 \; \text{on} \; \partial\Omega_u \},
         \label{eq:TestSpaceU}
      \end{equation}
      \begin{equation}
          V_s = \{\phi \in H^1(\Omega) \}.
      \end{equation}
  One can obtain the following weak form: Find $\boldsymbol{u}\in H_u$ and $s\in H_s$ such that for all $\boldsymbol{v}\in V_u$ and $\phi\in V_s$,
 \begin{subequations}
 \begin{align}
     a_{\text{Disp}}(\boldsymbol{u},\boldsymbol{v}) = b_{\text{Disp}}(\boldsymbol{v}),\\
    a_{\text{PF}}(s,\phi) = b_{\text{PF}}(\phi),
 \end{align}
 \label{eq:WeakForm}
\end{subequations}
where
 \begin{subequations}
 \begin{align}
    a_{\text{Disp}}(\boldsymbol{u},\boldsymbol{v}) = \int_{\Omega} \epsilon(\boldsymbol{v}):\sigma(\boldsymbol{u})\, d{\Omega},\,\,\,\,\,\,\,\,\,\,\,\,    b_{\text{Disp}}(\boldsymbol{v}) = {\bf{0}},\\
    a_{\text{PF}}(s,\phi) = \int_{\Omega}\bigl(G_c\,l_s\,\nabla s \cdot \nabla \phi + 2s\,\phi\,\mathcal{H}({\mathcal{E}}) +   \frac{G_c}{l_s}\,s \,\phi\bigr)\,d\Omega, \,\,\,\,\,\,\,\,\,\,\,\,      b_{\text{PF}}(\phi) = \int_{\Omega}\frac{G_c}{l_s}\phi \,d{\Omega}.
 \end{align}
 \label{eq:WeakFormParts}
\end{subequations}
In the present study, a staggered scheme originally proposed by Miehe {\it{et al.}} \cite{miehe2010phase} is employed to solve for the unknown displacement vector and phase-field from the weak form defined by Eq. \eqref{eq:WeakForm} and Eq. \eqref{eq:WeakFormParts} using Gridap.
      
One of the salient features of Gridap is that one can directly use the weak form in Julia for the finite element implementation using Gridap. All the steps associated with the finite element simulations such as creating the mesh file, implementing the weak form, application of boundary conditions, solutions for the unknown field variables, and post-processing for the output files are described through a Julia code on numerical simulation of a test on a notched beam under symmetric three-point bending (see Section \ref{sec:SymThreePointBendingTest} for numerical results). One may readily apply the developed phase-field-based Julia code for simulating other brittle fracture problems with appropriate modifications. For reproducing the results presented in Section \ref{Sec:NumericalSimulation}, one needs to first load the following Julia packages in the script file which is presently written in a jupyter notebook.
\begin{lstlisting}[language=C,backgroundcolor=\color{gray!5!white},showstringspaces=false,breaklines=true,basicstyle=\ttfamily]
using GridapGmsh
using Gridap
using Gridap.Geometry
using Gridap.TensorValues
using PyPlot
\end{lstlisting}
One can define the input parameters associated with the elastic and fracture material properties of the notched beam by writing the following lines.
\begin{lstlisting}[language=C,backgroundcolor=\color{gray!5!white},showstringspaces=false,breaklines=true,basicstyle=\ttfamily]
const E_mat = 20.8
const $\nu$_mat = 0.3

const Gc = 5e-4
const ls = 0.03
const $\eta$ = 1e-15
\end{lstlisting}
For the finite element simulations, one needs to have the discretization of the computational domain which can be generated by writing a mesh file in Julia (see \ref{Sec:AppendixFEmeshSymTPBtest}) and load that mesh file to build an instance of ``DiscreteModel" by the following lines. 
\begin{lstlisting}[language=C,backgroundcolor=\color{gray!5!white},showstringspaces=false,breaklines=true,basicstyle=\ttfamily]
model = GmshDiscreteModel("BeamWithNotchSymThreePtBending.msh")
writevtk(model,"BeamWithNotchSymThreePtBending")
\end{lstlisting}
For an isotropic material, the constitutive tensor $\mathbb{C}$ can be defined for two-dimensional plane stress and plane strain problems by writing the following function.
\begin{lstlisting}[language=C,backgroundcolor=\color{gray!5!white},showstringspaces=false,breaklines=true,basicstyle=\ttfamily]
function ElasFourthOrderConstTensor($E$,$\nu$,PlanarState)
    # 1 for Plane Stress and 2 Plane Strain Condition 
  if PlanarState == 1
      C1111 = $E$/(1-$\nu$*$\nu$)
      C1122 = ($\nu$*$E$)/(1-$\nu$*$\nu$)
      C1112 = 0.0
      C2222 = $E$/(1-$\nu$*$\nu$)
      C2212 = 0.0
      C1212 = $E$/(2*(1+$\nu$))
  elseif PlanarState == 2
      C1111 = ($E$*(1-$\nu$*$\nu$))/((1+$\nu$)*(1-$\nu$-2*$\nu$*$\nu$))
      C1122 = ($\nu$*$E$)/(1-$\nu$-2*$\nu$*$\nu$)
      C1112 = 0.0
      C2222 = ($E$*(1-$\nu$))/(1-$\nu$-2*$\nu$*$\nu$)
      C2212 = 0.0
      C1212 = $E$/(2*(1+$\nu$))
  end
  C_ten = SymFourthOrderTensorValue(C1111,C1112,C1122,C1112,C1212,C2212,C1122,C2212,C2222)
    return  C_ten
end
\end{lstlisting}
In the present study, plane strain condition is assumed for which the constitutive tensor is computed by calling the above Julia function as given below.
\begin{lstlisting}
const C_mat = ElasFourthOrderConstTensor(E_mat,$\nu$_mat,2)
\end{lstlisting}
To satisfy the assumption that crack can not propagate under pure compression, stress and strain tensors are decomposed into a volumetric and a deviatoric part by introducing the projection operators $\mathbb{P}_{\text{vol}}$ and $\mathbb{P}_{\text{dev}}$, respectively, which are defined by the following lines.
\begin{lstlisting}
I2 = SymTensorValue{2,Float64}(1.0,0.0,1.0)
I4 = I2$\otimes$I2
I4_sym = one(SymFourthOrderTensorValue{2,Float64})
P_vol = (1.0/3)*I4
P_dev = I4_sym - P_vol
\end{lstlisting}
To express the stress tensor in terms of kinematics variables i.e. $\boldsymbol{\sigma}=\mathbb{C}_{mod}\,\boldsymbol{\epsilon}$, where the expression of $\mathbb{C}_{mod}$ is given by Eq. \eqref{eq:CtensorMod}, the following function is defined in Julia. 
\begin{lstlisting}
function $\sigma$fun($\varepsilon$,$\varepsilon$_in,s_in)
    $\sigma$_elas = C_mat$\odot$$\varepsilon$
    if tr($\varepsilon$_in) >= 0
        $\sigma$ = (s_in ^2+$\eta$)*$\sigma$_elas
    elseif  tr($\varepsilon$_in) < 0
        $\sigma$ = (s_in ^2+$\eta$)*P_dev $\odot$ $\sigma$_elas + P_vol $\odot$ $\sigma$_elas
    end
    return  $\sigma$
end
\end{lstlisting}
One can determine the so-called positive part of elastic free energy,  $\psi^{\text{elas}}_+\left(\boldsymbol{\epsilon}\right)$ which is given by Eq. \eqref{eq:PosPartElastEnerg}, by writing the following function in Julia.
\begin{lstlisting}
function $\psi$Pos($\varepsilon$_in)
   $\sigma$_elas = C_mat$\odot$$\varepsilon$_in
    if tr($\varepsilon$_in) >= 0
        $\psi$Plus = 0.5*($\varepsilon$_in $\odot$ $\sigma$_elas)
    elseif  tr($\varepsilon$_in) < 0
        $\psi$Plus = 0.5*((P_dev $\odot$ $\sigma$_elas)$\odot$(P_dev$\odot$$\varepsilon$_in))
    end
    return $\psi$Plus
end
\end{lstlisting}
One needs to generate a discrete approximation of the finite element test and trial spaces of the problem on the discretized computational domain. Approximation of the finite element spaces associated with the phase field variable can be done by the following lines in Julia.
\begin{lstlisting}
order = 1
reffe_PF = ReferenceFE(lagrangian,Float64,order)
V0_PF = TestFESpace(model,reffe_PF;
  conformity=:H1)
U_PF = TrialFESpace(V0_PF)
sh = zero(V0_PF)
\end{lstlisting}
Similarly, one can generate the approximation of the finite element spaces associated with the displacement variable by writing the following lines in Julia.
\begin{lstlisting}
reffe_Disp = ReferenceFE(lagrangian,VectorValue{2,Float64},order)
    V0_Disp = TestFESpace(model,reffe_Disp;
     conformity=:H1,
    dirichlet_tags=["LeftSupport","RightSupport","LoadLine"],
    dirichlet_masks=[(false,true), (true,true), (false,true)])
uh = zero(V0_Disp)
\end{lstlisting}
To compute the integrals in the weak form given by Eq. \eqref{eq:WeakFormParts} numerically, one needs to define an integration mesh along with an integration rule (for example, Gauss quadrature) in each of the cells in the triangulation. Using Gridap, one can easily define the integration mesh and the corresponding Lebesgue measure by using the built-in functions ``Triangulation" and ``Measure", respectively. For instance, one can use the following lines for integrating the weak form given by Eq. \eqref{eq:WeakFormParts} defined on the domain $\Omega$
using a quadrature rule of degree two times the order of interpolation in the cells of the triangulation.
\begin{lstlisting}
degree = 2*order
$\Omega$ = Triangulation(model)
d$\Omega$ = Measure($\Omega$,degree)
\end{lstlisting}
One can determine the applied load on a part of the boundary of the domain by determining boundary integral using the following built-in functions available in Gridap.
\begin{lstlisting}
labels = get_face_labeling(model)
LoadTagId = get_tag_from_name(labels,"LoadLine")
$\Gamma$_Load = BoundaryTriangulation(model,tags = LoadTagId)
d$\Gamma$_Load = Measure($\Gamma$_Load,degree)
n_$\Gamma$_Load = get_normal_vector($\Gamma$_Load)
\end{lstlisting}
To find the values of variables that are defined by using the built-in function ``CellState" at the Gauss points, one need to use the ``project" function as defined below.
\begin{lstlisting}
function project($q$,model,d$\Omega$,order)
  reffe = ReferenceFE(lagrangian,Float64,order)
  $\text{V}$ = FESpace(model,reffe,conformity=:L2)
  a($u$,$v$) = $\int$($u$*$v$)*d$\Omega$
  b($v$) = $\int$($v$*$q$)*d$\Omega$
  op = AffineFEOperator(a,b,$\text{V}$,$\text{V}$)
  qh = solve(op)
  return qh
end
\end{lstlisting}
In the present study, a staggered scheme is used to update the solution from the pseudo time $t_n$ to $t_{n+1}$ \cite{miehe2010phase}.       Given the displacement vector, phase-field and the history function at the time $t_n$, one can update the phase-field at the time $t_{n+1}$ by using the following function.
\begin{lstlisting}
function  stepPhaseField(uh_in,$\Psi$PlusPrev_in)
  a_PF($s$,$\phi$) = $\int$( Gc*ls*$\nabla(\phi)$$\cdot \nabla(s)$ + 2*$\Psi$PlusPrev_in*$s$*$\phi$ + (Gc/ls)*$s$*$\phi$ )*d$\Omega$
  b_PF($\phi$) = $\int$( (Gc/ls)*$\phi$)*d$\Omega$
  op_PF = AffineFEOperator(a_PF,b_PF,U_PF,V0_PF)
  sh_out = solve(op_PF)  
    return sh_out
end
\end{lstlisting}
Using the values of displacement vector and the history function at time $t_n$, and the computed value of phase-field at time $t_{n+1}$, one can update the displacement vector at time $t_{n+1}$ by calling a function as given below.
\begin{lstlisting}
 function   stepDisp(uh_in,sh_in,vApp)
    uApp1($x$) = VectorValue(0.0,0.0)
    uApp2($x$) = VectorValue(0.0,0.0)
    uApp3($x$) = VectorValue(0.0,-vApp)
    U_Disp = TrialFESpace(V0_Disp,[uApp1,uApp2,uApp3])

    a_Disp($u$,$v$) = $\int$( ($\varepsilon(v)$$\odot$ ($\sigma$fun$\circ$($\varepsilon(u)$,$\varepsilon$(uh_in),sh_in)) ) )*d$\Omega$
    b_Disp($v$) = 0.0
    op_Disp = AffineFEOperator(a_Disp,b_Disp,U_Disp,V0_Disp)
    uh_out = solve(op_Disp)
    return uh_out
end
\end{lstlisting}
Once both displacement and phase field are determined at time $t_{n+1}$, one can update the energy history function at time $t_{n+1}$ by defining it as
\begin{equation}
 \mathcal{H}_{n+1} = 
 \begin{cases}
    \mathcal{E}_{n+1}\,\,\text{for}\,\, \mathcal{E}_{n+1} > \mathcal{E}_{n}& \\
    \mathcal{E}_{n} \,\,\text{otherwise},
  \end{cases}
 \end{equation}
where the history function $\mathcal{H}(\mathcal{E})$ of elastic free energy $\mathcal{E} =\psi^{\text{elas}}_+(\boldsymbol{\epsilon})$ need to be defined in Julia (see Section \ref{sec:BVPdef}), which can be achieved by writing the following lines.
\begin{lstlisting}
function new_EnergyState($\psi$PlusPrev_in,$\psi$hPos_in)
  $\psi$Plus_in = $\psi$hPos_in
  if $\psi$Plus_in >= $\psi$PlusPrev_in
    $\psi$Plus_out = $\psi$Plus_in
  else
    $\psi$Plus_out = $\psi$PlusPrev_in
  end
  true,$\psi$Plus_out
end
\end{lstlisting}
Finally, one can simulate the brittle fracture in the notched beam by applying a monotonic displacement control loading and solve for the unknown displacement and phase-field at each loading step by writing the main routine in Julia that uses the above-defined functions as listed below.  
\begin{lstlisting}[language=C,backgroundcolor=\color{gray!5!white},showstringspaces=false,breaklines=true,basicstyle=\ttfamily]
vApp = 0
delv = 1e-3
const vAppMax = 0.1
innerMax = 10
count = 0
Load = Float64[]
Displacement = Float64[]
push!(Load, 0.0)
push!(Displacement, 0.0)
sPrev = CellState(1.0,d$\Omega$)
sh = project(sPrev,model,d$\Omega$,order)
$\psi$PlusPrev = CellState(0.0,d$\Omega$)

while vApp .< vAppMax 
    count = count .+ 1
    if vApp >= 3e-2
        delv = 1e-4
    end
    vApp = vApp .+ delv
    print("\n Entering  displacemtent  step :", float(vApp))
   for inner = 1:innerMax
        $\psi$hPlusPrev = project($\psi$PlusPrev,model,d$\Omega$,order)
        
        RelErr = abs(sum($\int$( Gc*ls*$\nabla$(sh)$\cdot$$\nabla$(sh) + 2*$\psi$hPlusPrev*sh*sh + (Gc/ls)*sh*sh)*d$\Omega$-$\int$( (Gc/ls)*sh)*d$\Omega$))/abs(sum($\int$( (Gc/ls)*sh)*d$\Omega$))
        
        sh = stepPhaseField(uh,$\psi$hPlusPrev) 
        uh = stepDisp(uh,sh,vApp)
        
        $\psi$hPos_in = $\psi$Pos$\circ$($\varepsilon$(uh))    
        update_state!(new_EnergyState,$\psi$PlusPrev,$\psi$hPos_in)
        
        if  RelErr < 1e-8
            break 
        end      
    end
    Node_Force = sum($\int$(n_$\Gamma$_Load$\cdot$($\sigma$fun$\circ$($\varepsilon$(uh),$\varepsilon$(uh),sh))) *d$\Gamma$_Load)
    push!(Load, -Node_Force[2])
    push!(Displacement, vApp)
    end
end 
\end{lstlisting}
One can create output files at each loading step that can be viewed in ParaView. For instance, one can create a ``.vtu" file to save data for the solution of the displacement vector and phase-field at each loading step by using the following lines in Julia. 
\begin{lstlisting}
writevtk($\Omega$,"results_SymThreePtBendingTest",cellfields
= ["uh"=>uh,"s"=>sh])
\end{lstlisting}
One can generate the load-displacement curve by using the plot command as given below.
\begin{lstlisting}
plot(Displacement,Load)
\end{lstlisting}

\section{Numerical Simulations}
\label{Sec:NumericalSimulation}
Proposed phase-field-based Julia codes are validated against a test on a notched beam under symmetric three-point bending and a set of tests on a notched beam with three holes under asymmetric three-point bending. The effect of length scale parameter value on the fracture response is well understood and hence not repeated here. To validate the proposed open-source implementation, one particular value of the length scale parameter $l_s$, which is typically mentioned in the literature for the given problem, is taken. For the numerical simulation, a non-uniform finite element mesh with a finer mesh (length of the largest side of triangular elements is less than half of $l_s$ value) in regions where cracks may propagate is used. For finite element mesh generation of the notched beams used for symmetric and the asymmetric three-point bending test, Julia codes are provided in \ref{Sec:AppendixFEmeshSymTPBtest} and \ref{Sec:AppendixFEmeshAsymTPBtest},  respectively.

\subsection{Symmetric three point bending test}
\label{sec:SymThreePointBendingTest}
\begin{figure}[ht!]
  \subfloat[]{
	\begin{minipage}[c][0.52\width]{0.52\textwidth}
	   \centering
	 \includegraphics[width=\textwidth]{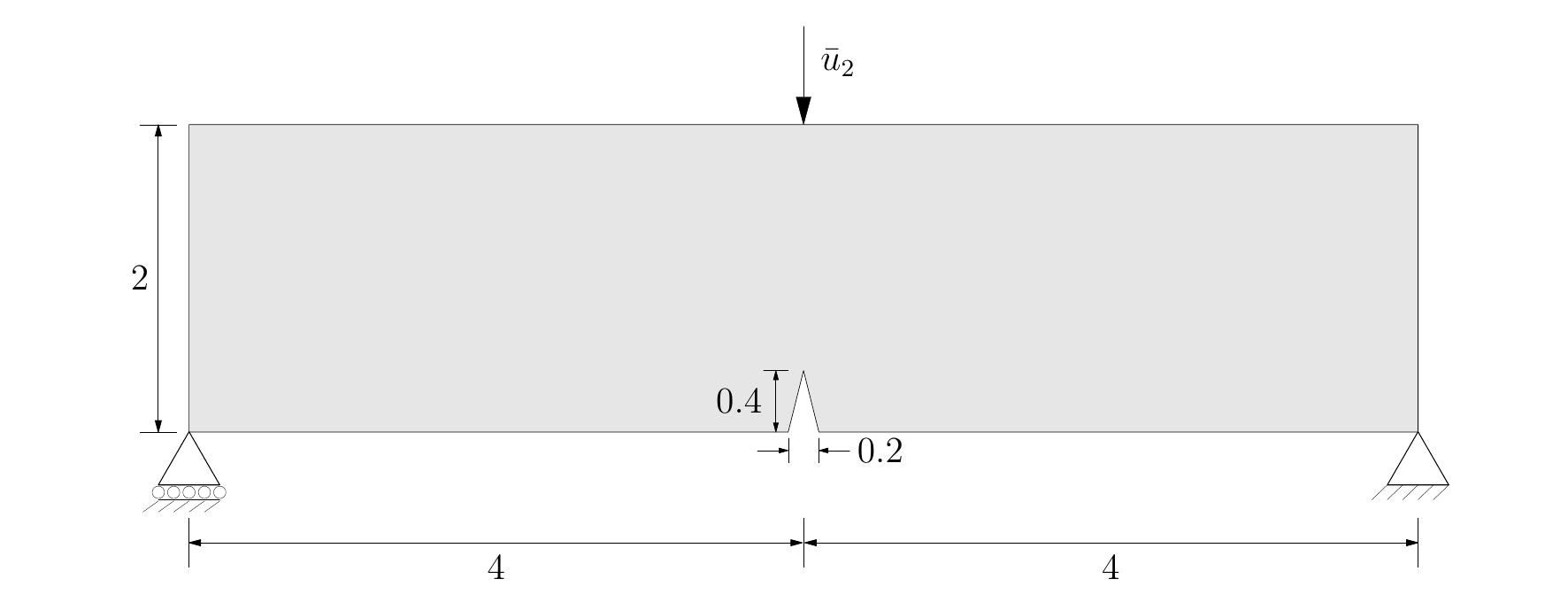}
	\end{minipage}}
  \subfloat[]{
	\begin{minipage}[c][0.62\width]{0.46\textwidth}
	  \centering
	 \includegraphics[width=\textwidth]{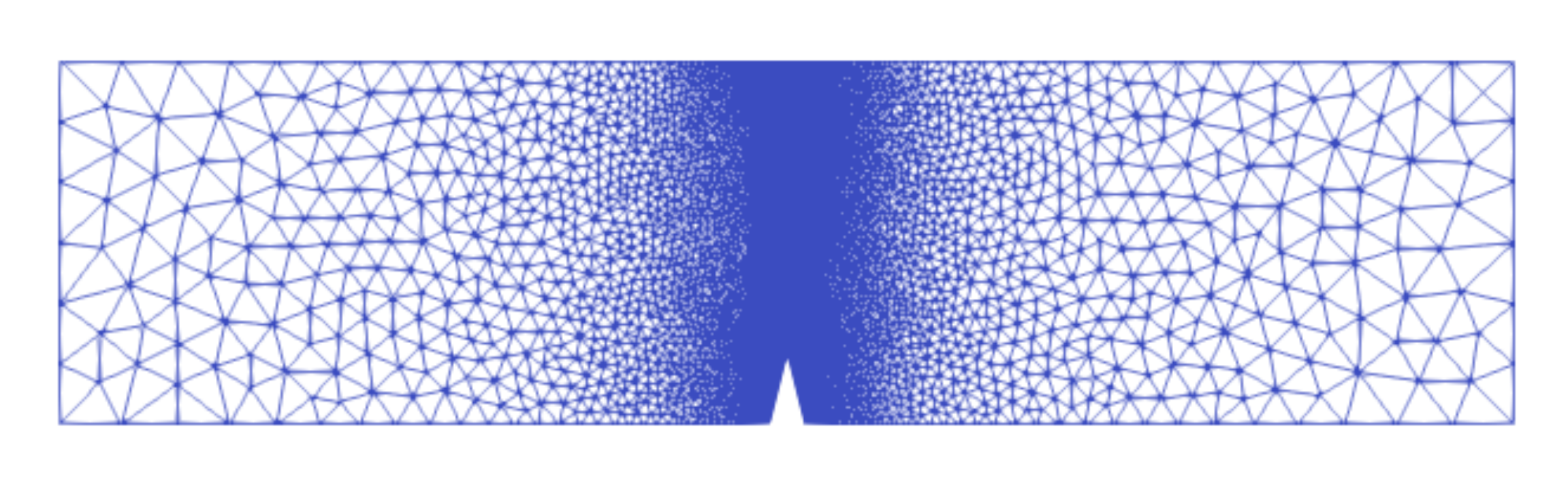}
	\end{minipage}}
\caption{Three point bending test set-up and a finite element mesh for the geometry of the notched beam (all dimensions are in millimeters (mm)). Sub-figure (a) shows the geometry and boundary conditions for the test. Sub-figure (b) shows the mesh using triangular elements for the finite element simulation.}
\label{fig:SymTPBtest}
\end{figure}

\begin{figure}[ht!]
     \centering
 \begin{subfigure}[b]{\textwidth}
         \centering
 \includegraphics[width=\textwidth]{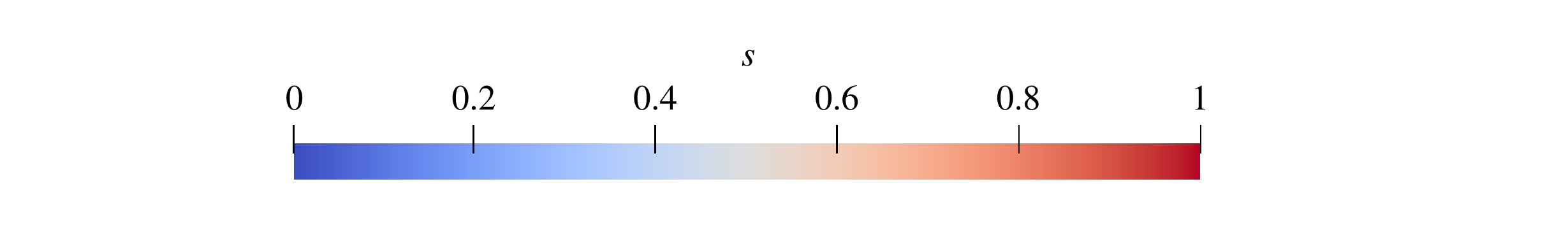}
     \end{subfigure}
     \begin{subfigure}[b]{0.49\textwidth}
         \centering
 \includegraphics[width=\textwidth]{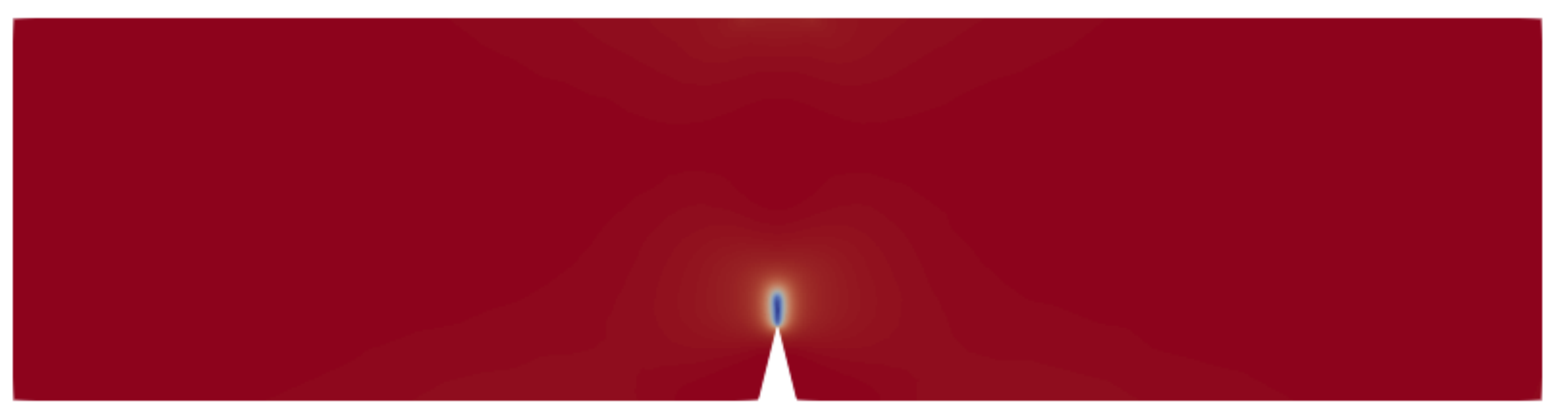}
         \caption{}
     \end{subfigure}
     \begin{subfigure}[b]{0.49\textwidth}
    \centering
 	 \includegraphics[width=\textwidth]{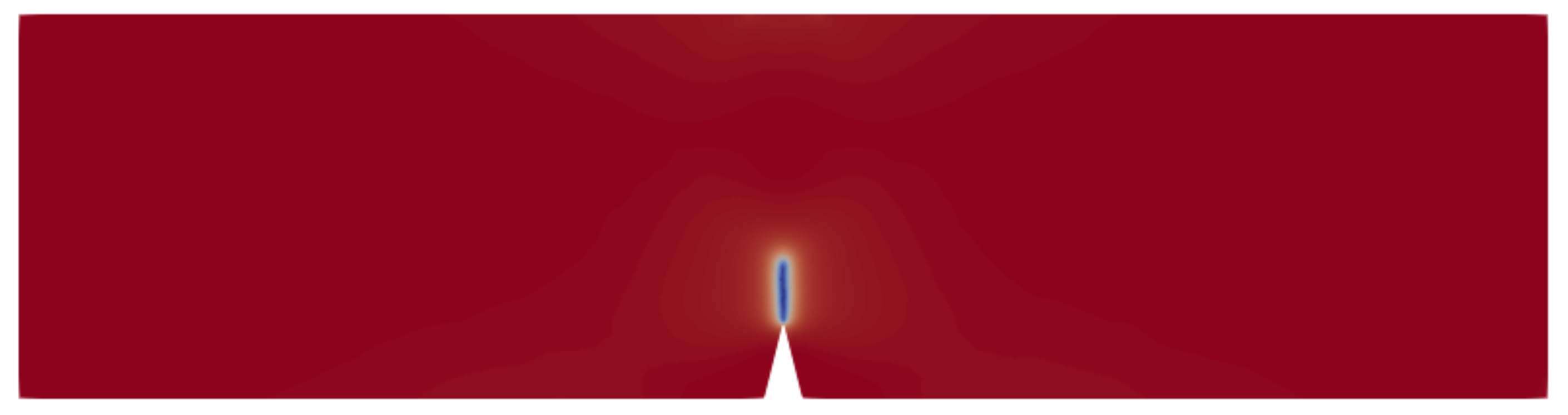}
         \caption{}
     \end{subfigure}
 \begin{subfigure}[b]{0.49\textwidth}
    \centering
 	 \includegraphics[width=\textwidth]{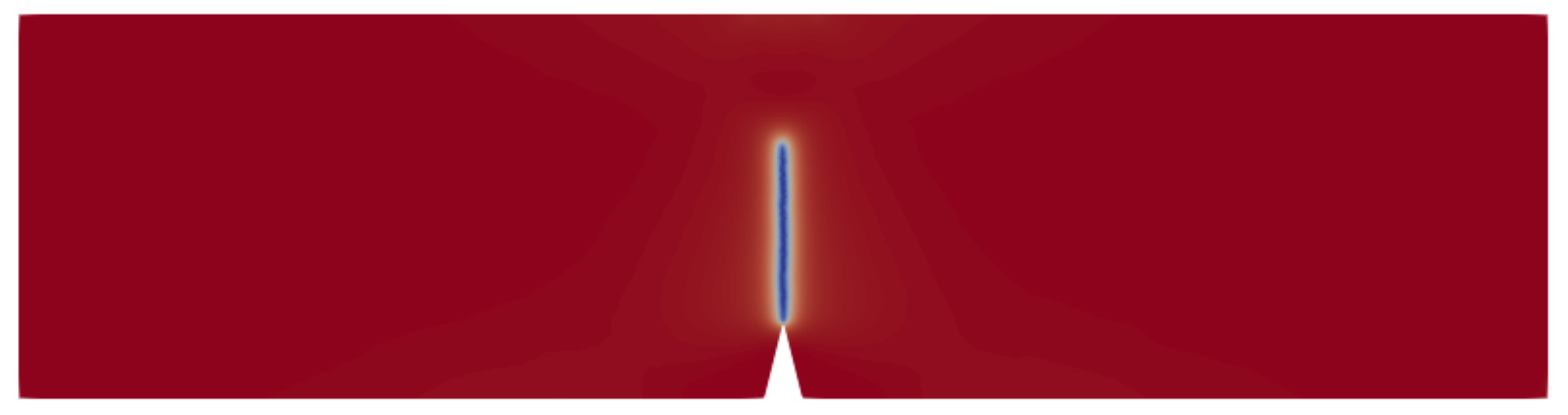}
         \caption{}
     \end{subfigure}
 \begin{subfigure}[b]{0.49\textwidth}
    \centering
 	 \includegraphics[width=\textwidth]{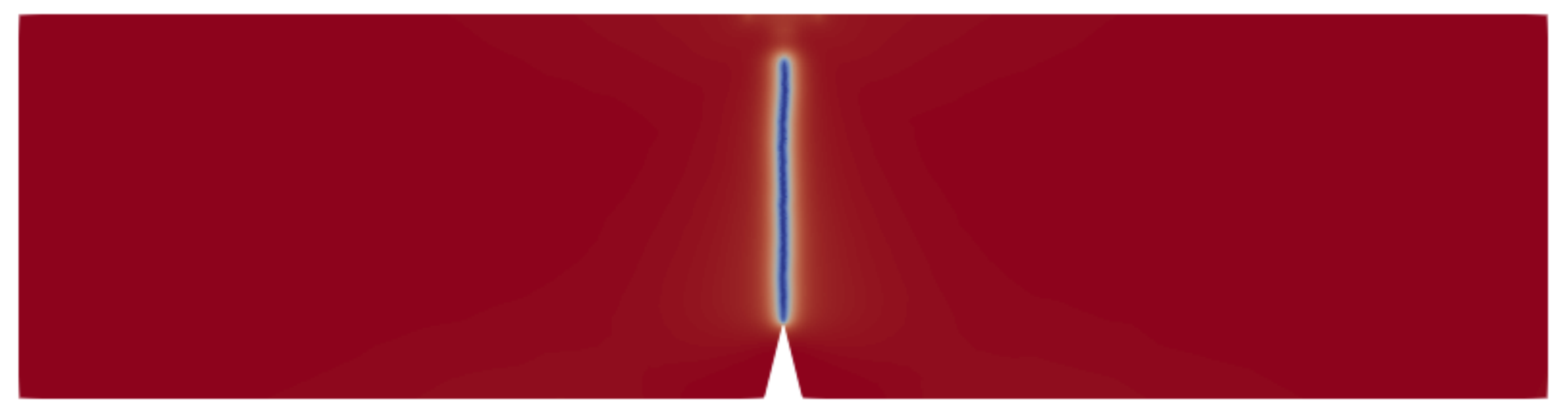}
         \caption{}
     \end{subfigure}
\caption{Damage  profiles for a notched beam under symmetric three point bending at different  deformation stages shown in sub-figures at  applied  displacement  (a) 0.0405 mm (b) 0.041 mm (c) 0.05 mm and (d) 0.1 mm.}
\label{fig:DamageProfileSymTPBtest}
\end{figure}

\begin{figure}[ht!]
{\begin{minipage}{\textwidth}
\centering
	 \includegraphics[width=0.75\textwidth]{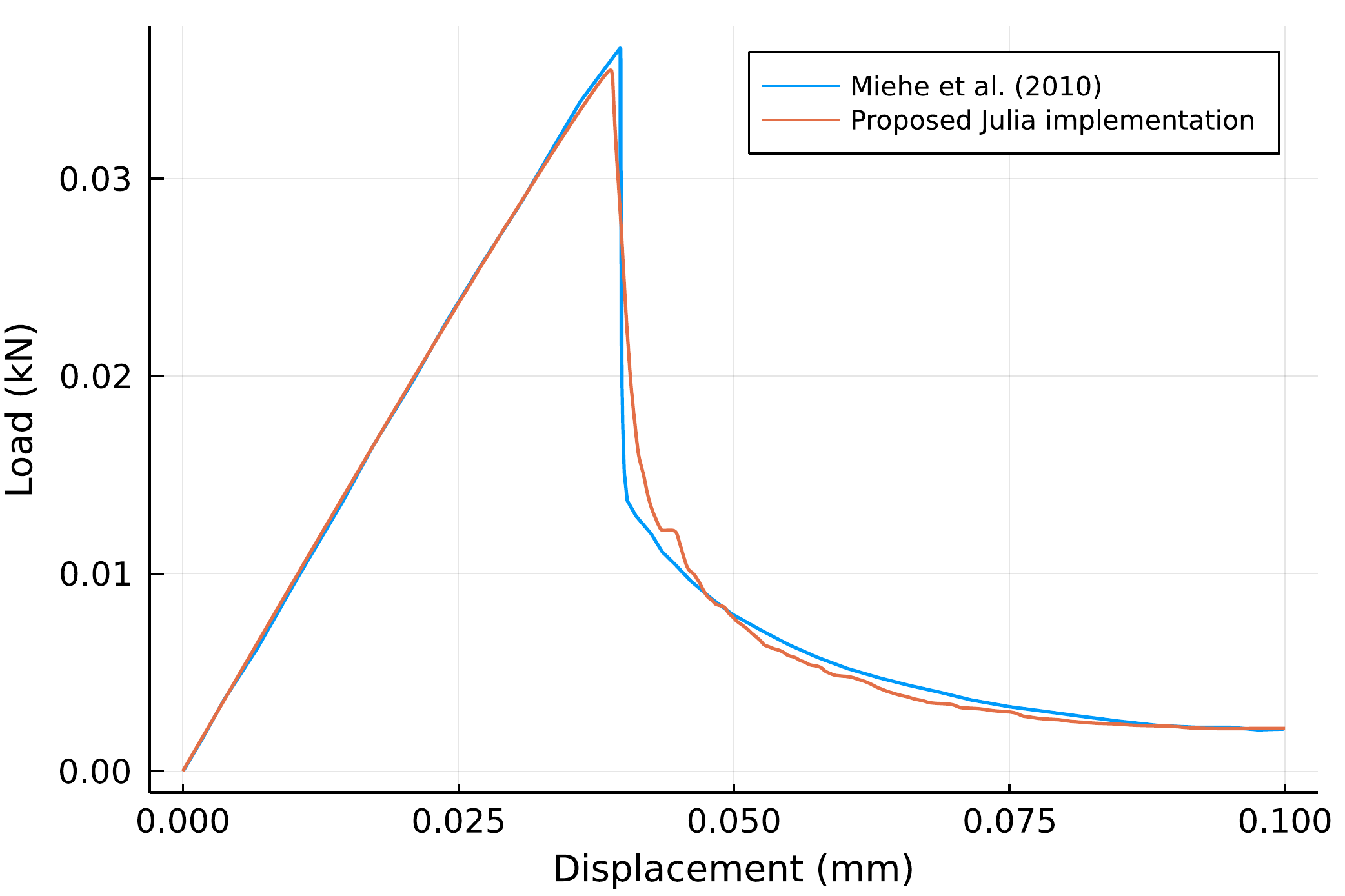}
	\end{minipage}}
\caption{Load-deﬂection curves for a symmetric three point bending test.}
\label{fig:LoadDispSymTPBtest}
\end{figure}

Modeling of brittle fracture in a simply supported notched beam under symmetric three-point bending is one of the classical benchmark problems which has frequently been analyzed in the literature \cite{ambati2015review,miehe2010phase,miehe2007robust,wu2018length}. The three-point bending test set-up and a finite element mesh used for the simulation are demonstrated in Fig. \ref{fig:SymTPBtest}.
For the numerical simulation, material properties for the notched beam are taken as
$E = 2.08\times 10^{4} \,\text{MPa}$, $\nu = 0.3$, $G_c = 5.0\times 10^{-4} \,\text{kN}/\text{mm}$ and $l_s = 0.03 \,\text{mm}$. Displacement control loading (monotonic displacement $\bar{u}_2$ is applied in small increments $\Delta\bar{u}_2$) is considered and the damage profiles for the notched beam at different stages of applied displacement are presented in Fig. \ref{fig:DamageProfileSymTPBtest}.
 For applied displacement up to $\bar{u}_2 = 0.025\,\text{mm}$, monotonic increment of $\Delta\bar{u}_2 = 10^{-3}\,\text{mm}$ and from $\bar{u}_2 = 0.025\,\text{mm}$ to until failure (here $\bar{u}_2 = 0.1\,\text{mm}$) monotonic increment of $\Delta\bar{u}_2 = 10^{-4}\,\text{mm}$ is used. As can be seen from Fig. \ref{fig:LoadDispSymTPBtest}, load displacement curve using the proposed open source implementation matches quite well with the  results reported in the literature \cite{miehe2010phase}.

\subsection{Asymmetric notched three point bending test}

\begin{figure}[ht!]
  \subfloat[]{
	\begin{minipage}[c][0.52\width]{0.52\textwidth}
	   \centering
	 \includegraphics[width=\textwidth]{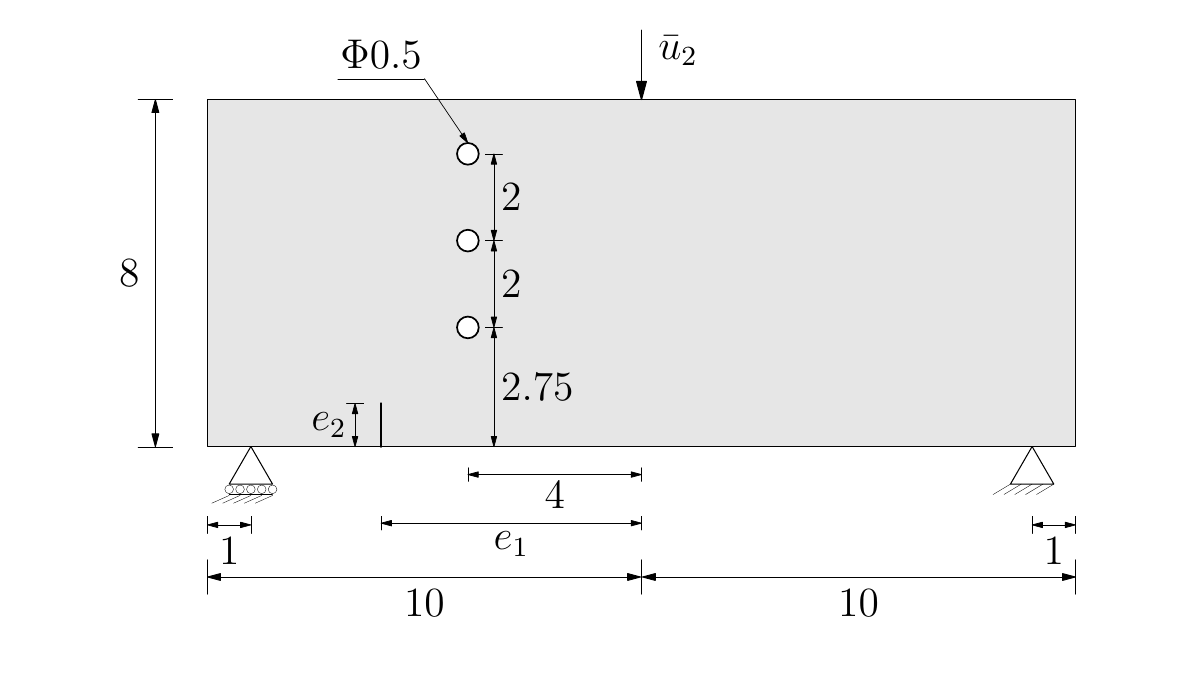}
	\end{minipage}}
  \subfloat[]{
	\begin{minipage}[c][0.83\width]{0.4\textwidth}
	  \centering
	 \includegraphics[width=\textwidth]{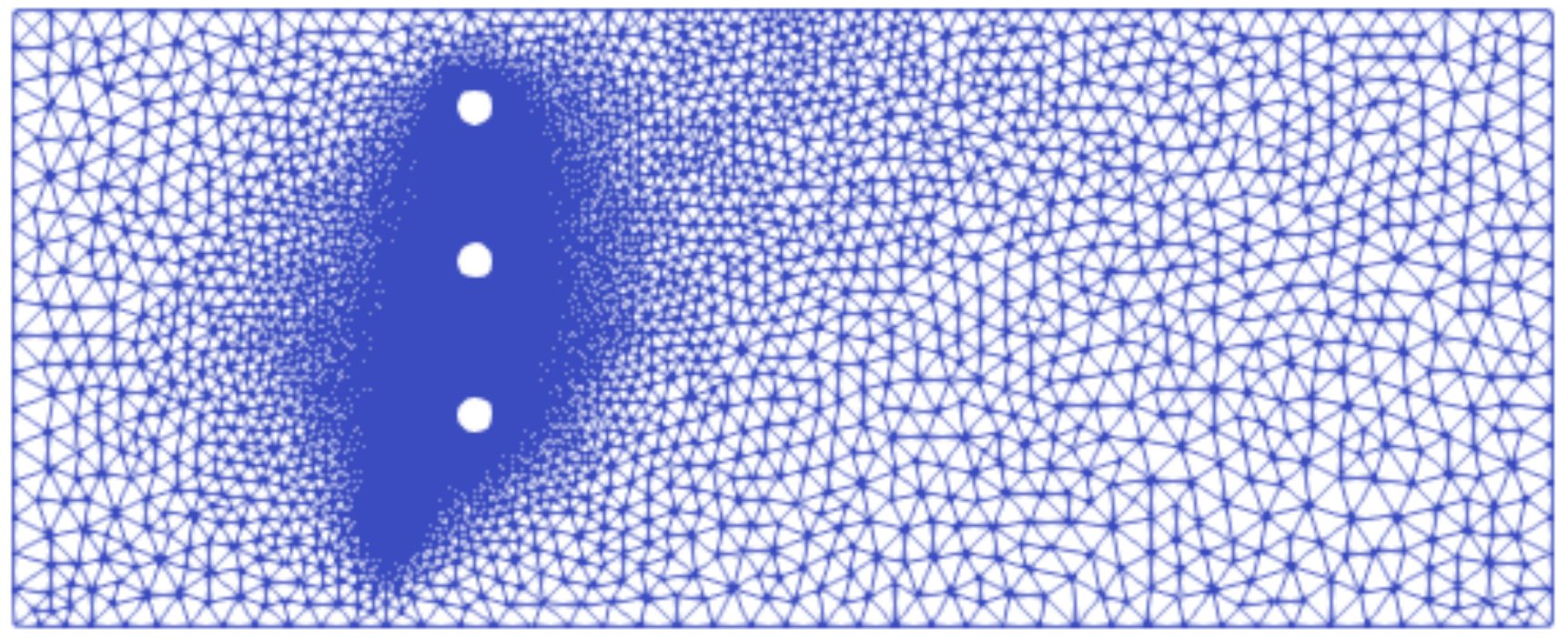}
	\end{minipage}}
\caption{Asymmetric three point bending test set-up and a finite element mesh for the geometry of the notched beam with three holes (all dimensions are in inches (in)). Sub-figure (a) shows the geometry and boundary conditions for the test. Sub-figure (b) shows the mesh using triangular elements for the finite element simulation.}
\label{fig:AsymTPBtestSetUp}
\end{figure}

\begin{figure}[ht!]
     \centering
 \begin{subfigure}[b]{\textwidth}
         \centering
 \includegraphics[width=\textwidth]{Figures/ColorbarForAsymThreePointBendingTest-eps-converted-to.pdf}
     \end{subfigure}
     \begin{subfigure}[b]{0.49\textwidth}
         \centering
 \includegraphics[width=\textwidth]{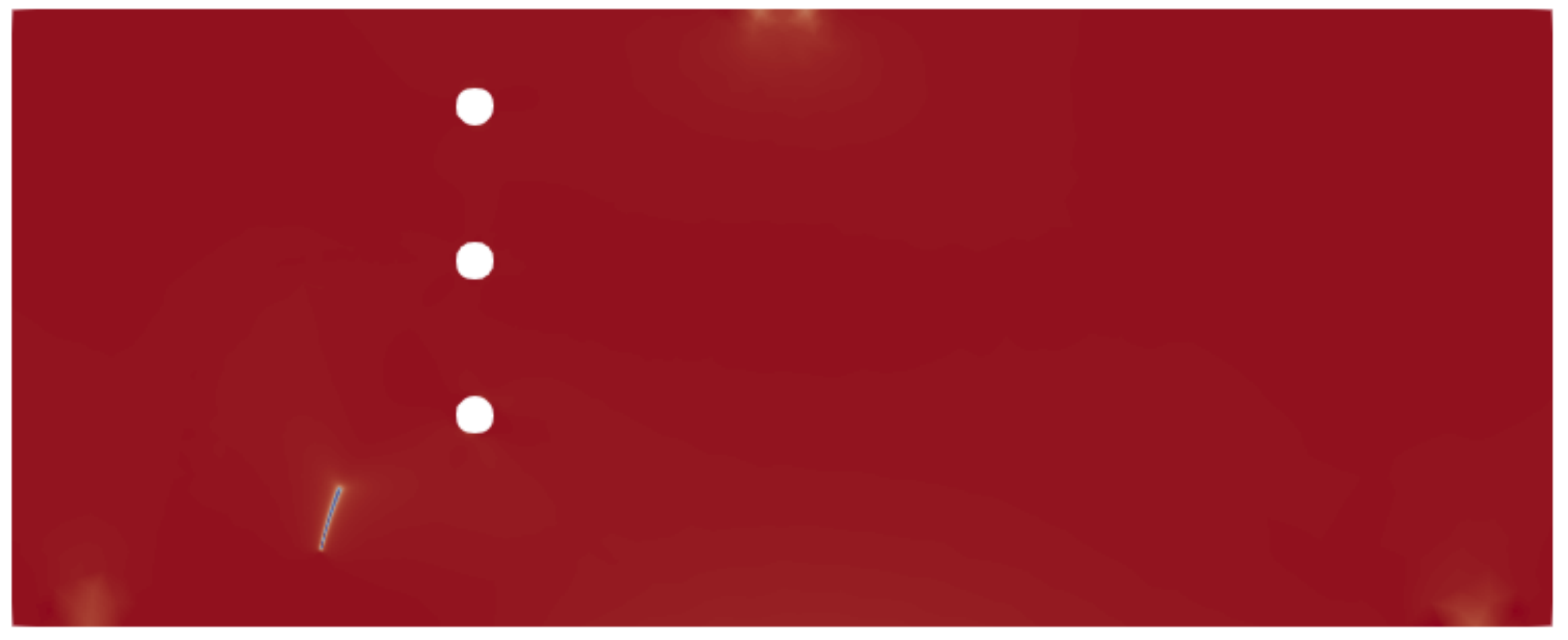}
         \caption{}
     \end{subfigure}
     \begin{subfigure}[b]{0.49\textwidth}
    \centering
 	 \includegraphics[width=\textwidth]{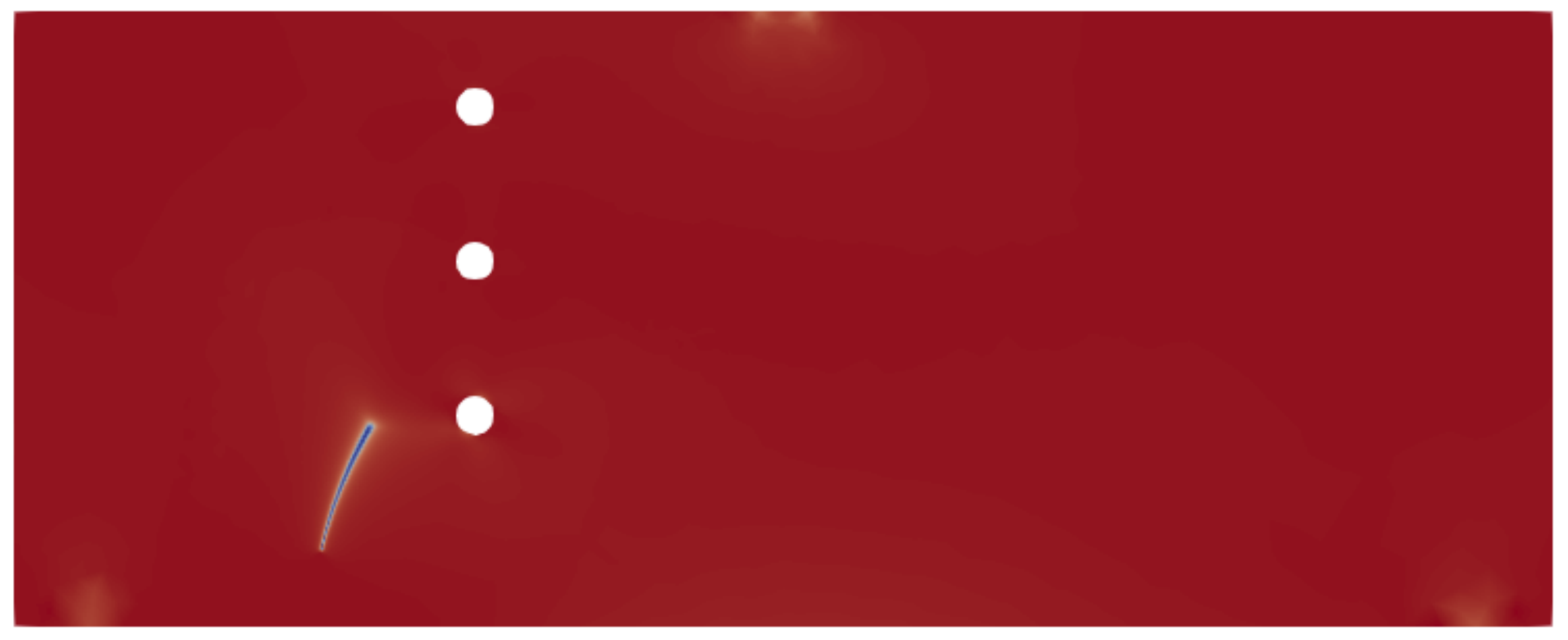}
         \caption{}
     \end{subfigure}
 \begin{subfigure}[b]{0.49\textwidth}
    \centering
 	 \includegraphics[width=\textwidth]{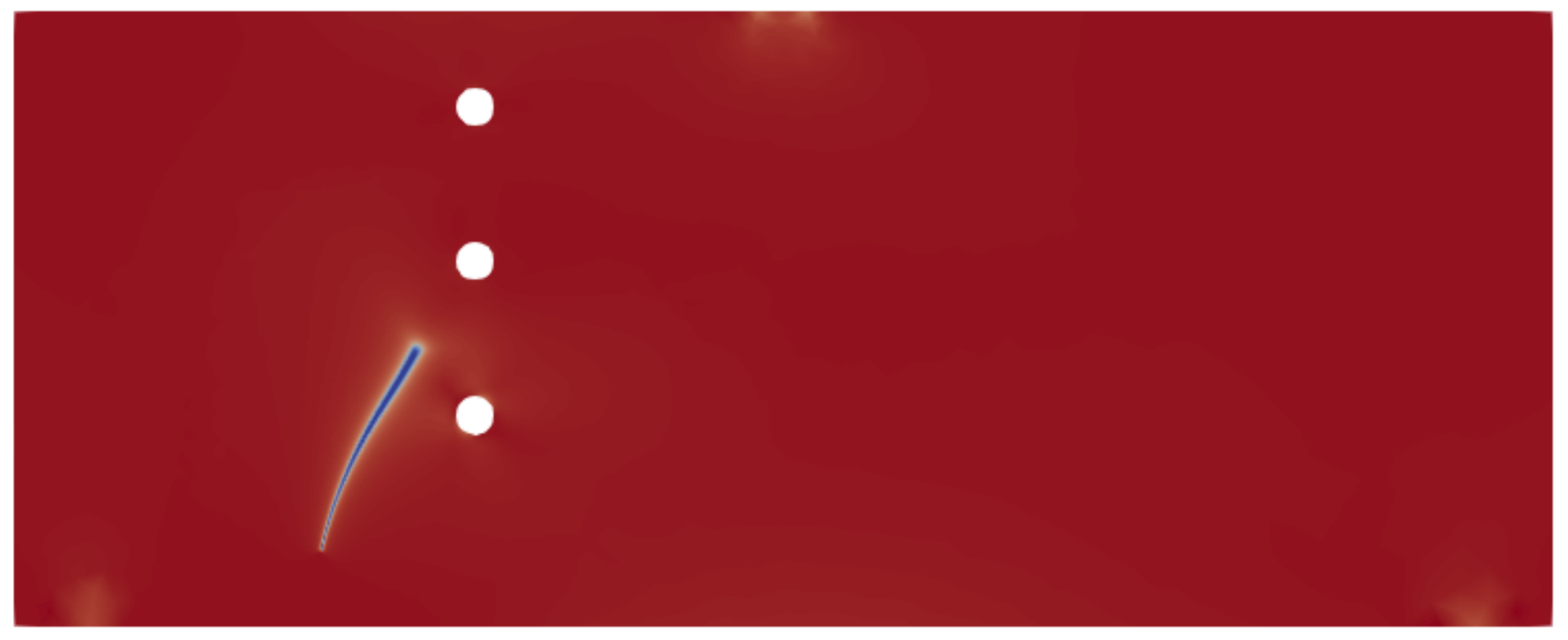}
         \caption{}
     \end{subfigure}
 \begin{subfigure}[b]{0.49\textwidth}
    \centering
 	 \includegraphics[width=\textwidth]{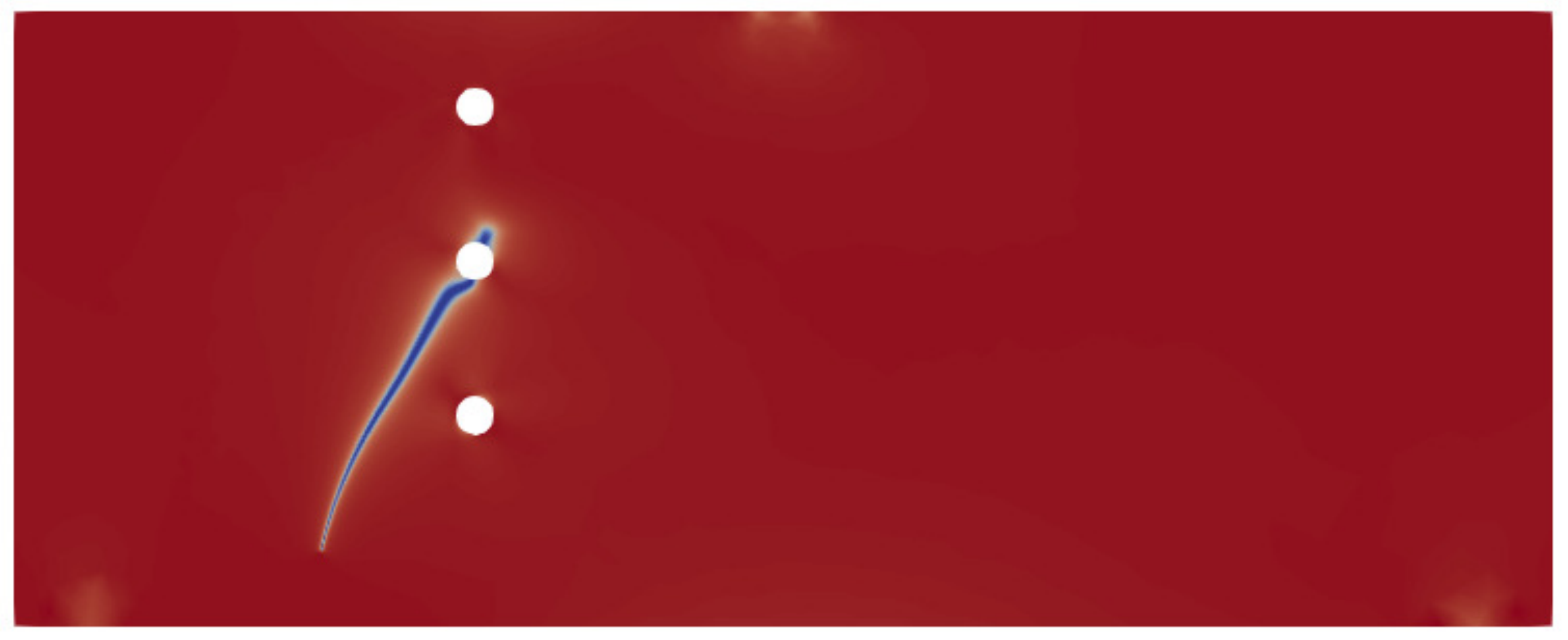}
         \caption{}
     \end{subfigure}
\caption{Damage  profiles for a beam with three holes and a pre-notch defined by $e_1 = 6\, \text{inch}$ and $e_2 = 1 \, \text{inch}$ under asymmetric three point bending at different  deformation  stages  shown  in  sub-figures  at  applied  displacement  (a) 0.055 inch (b) 0.0585 inch (c) 0.061 inch and (d) 0.0625 inch.}
\label{Fig:AsymTPBtestDamageProfileSpecimenA}
\end{figure}

\begin{figure}[ht!]
     \centering
     \begin{subfigure}[b]{0.35\textwidth}
         \centering
 \includegraphics[width=\textwidth]{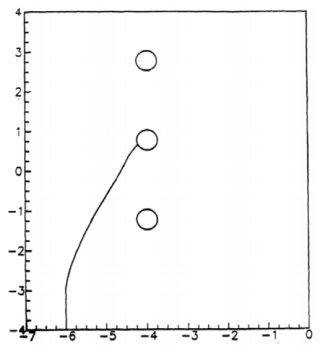}
         \caption{}
     \end{subfigure}
     \begin{subfigure}[b]{0.255\textwidth}
    \centering
 	 \includegraphics[width=\textwidth]{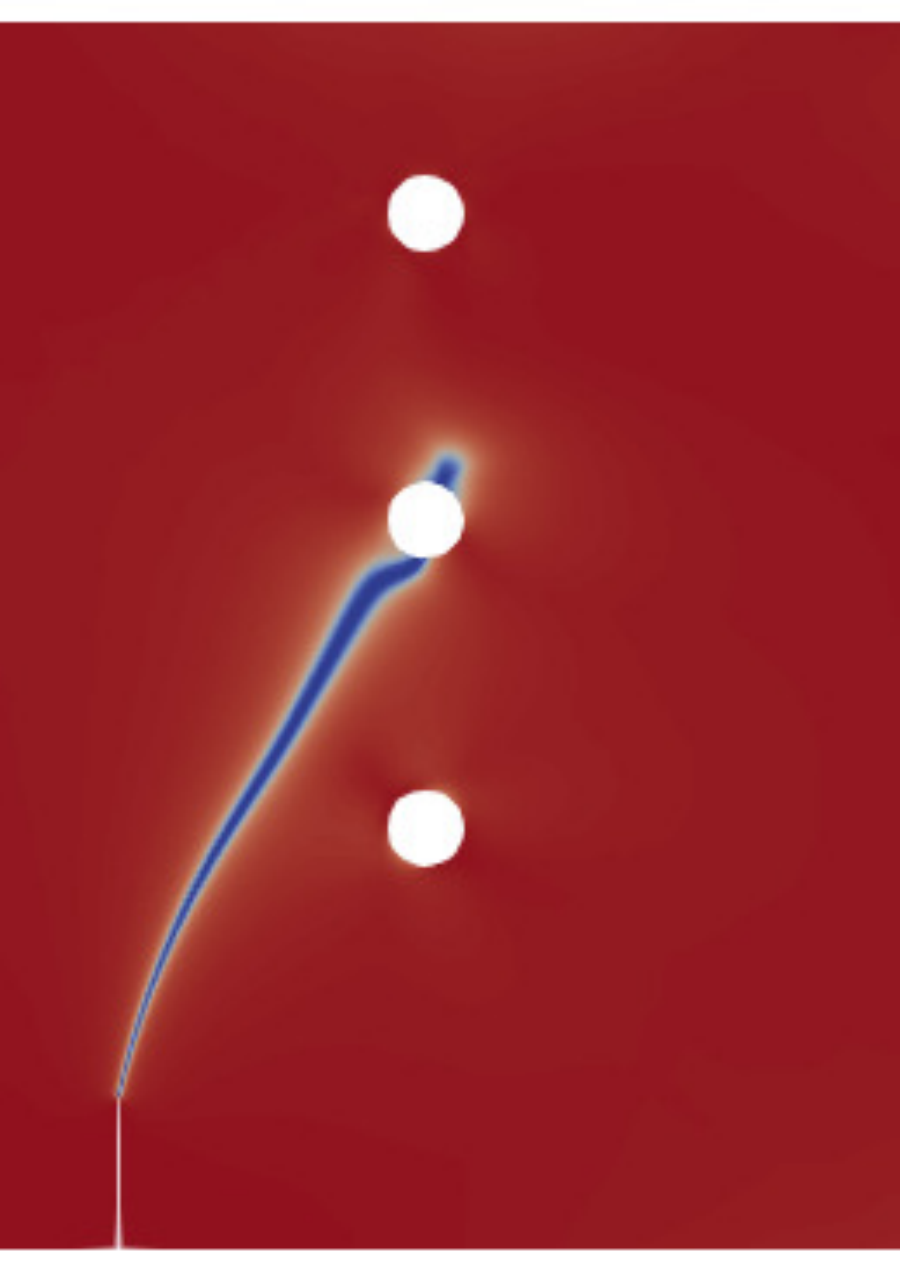}
         \caption{}
     \end{subfigure}
\caption{Comparison of the crack paths for the beam with three holes and a pre-notch defined by $e_1 = 6\, \text{inch}$ and $e_2 = 1 \, \text{inch}$ under asymmetric three point bending. Sub-figures (a) and (b) show crack paths for the experimentally observed \cite{ingraffea1990probabilistic} and the numerically predicted, respectively.}
\label{Fig:AsymTPBtestCrackPathSpecimenA}
\end{figure}

\begin{figure}[ht!]
     \centering
 \begin{subfigure}[b]{\textwidth}
         \centering
 \includegraphics[width=\textwidth]{Figures/ColorbarForAsymThreePointBendingTest-eps-converted-to.pdf}
     \end{subfigure}
     \begin{subfigure}[b]{0.49\textwidth}
         \centering
 \includegraphics[width=\textwidth]{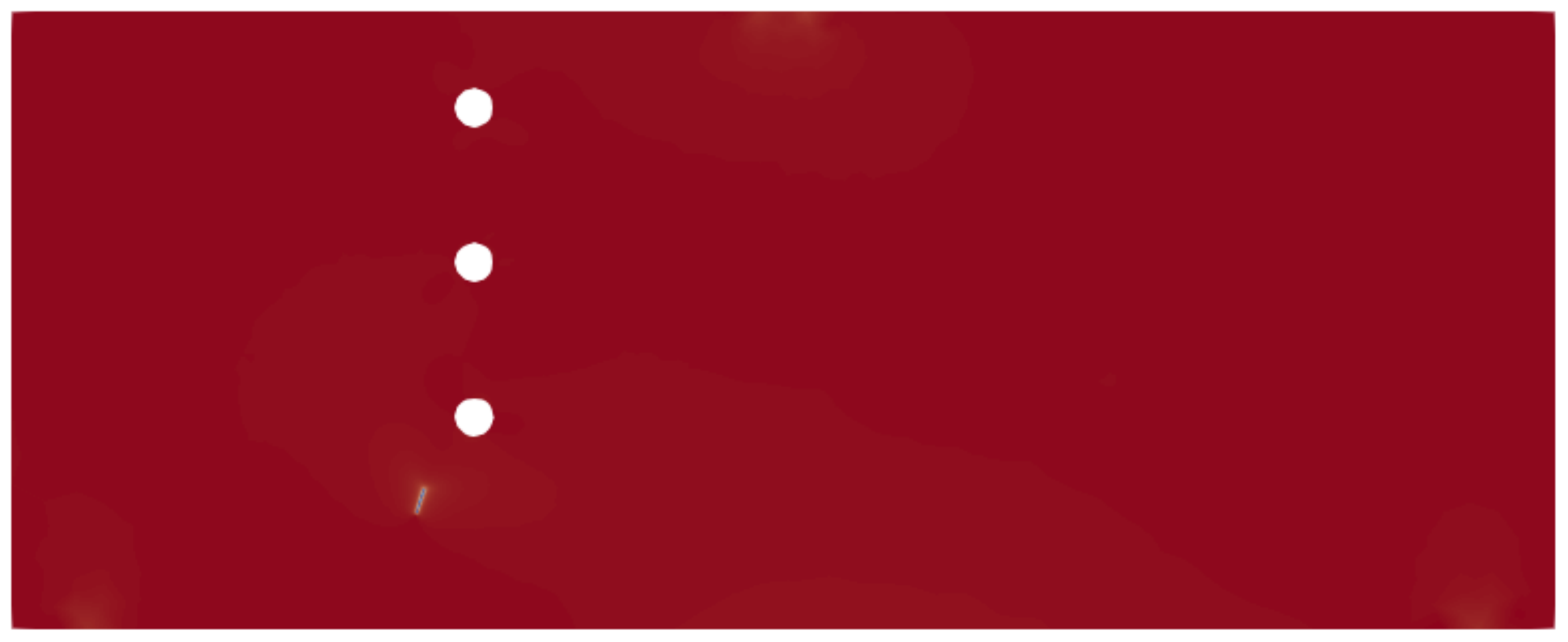}
         \caption{}
     \end{subfigure}
     \begin{subfigure}[b]{0.49\textwidth}
    \centering
 	 \includegraphics[width=\textwidth]{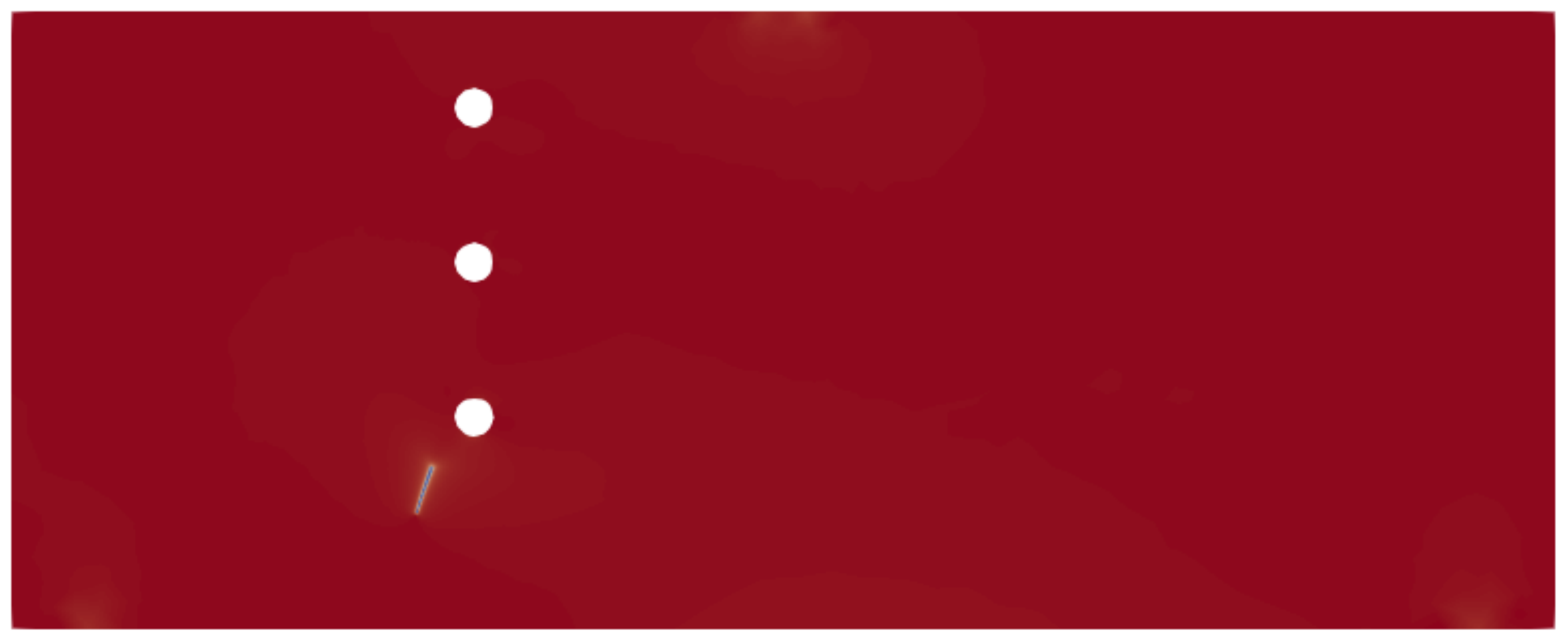}
         \caption{}
     \end{subfigure}
 \begin{subfigure}[b]{0.49\textwidth}
    \centering
 	 \includegraphics[width=\textwidth]{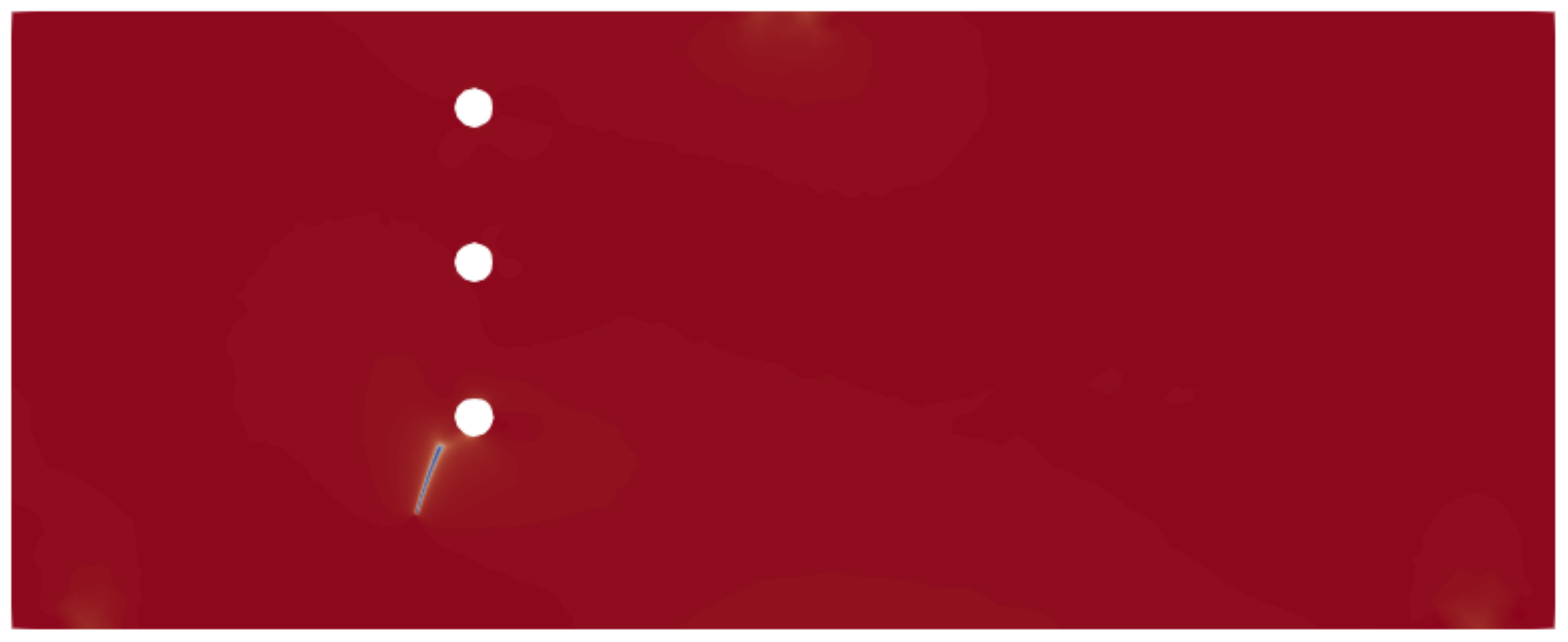}
         \caption{}
     \end{subfigure}
 \begin{subfigure}[b]{0.49\textwidth}
    \centering
 	 \includegraphics[width=\textwidth]{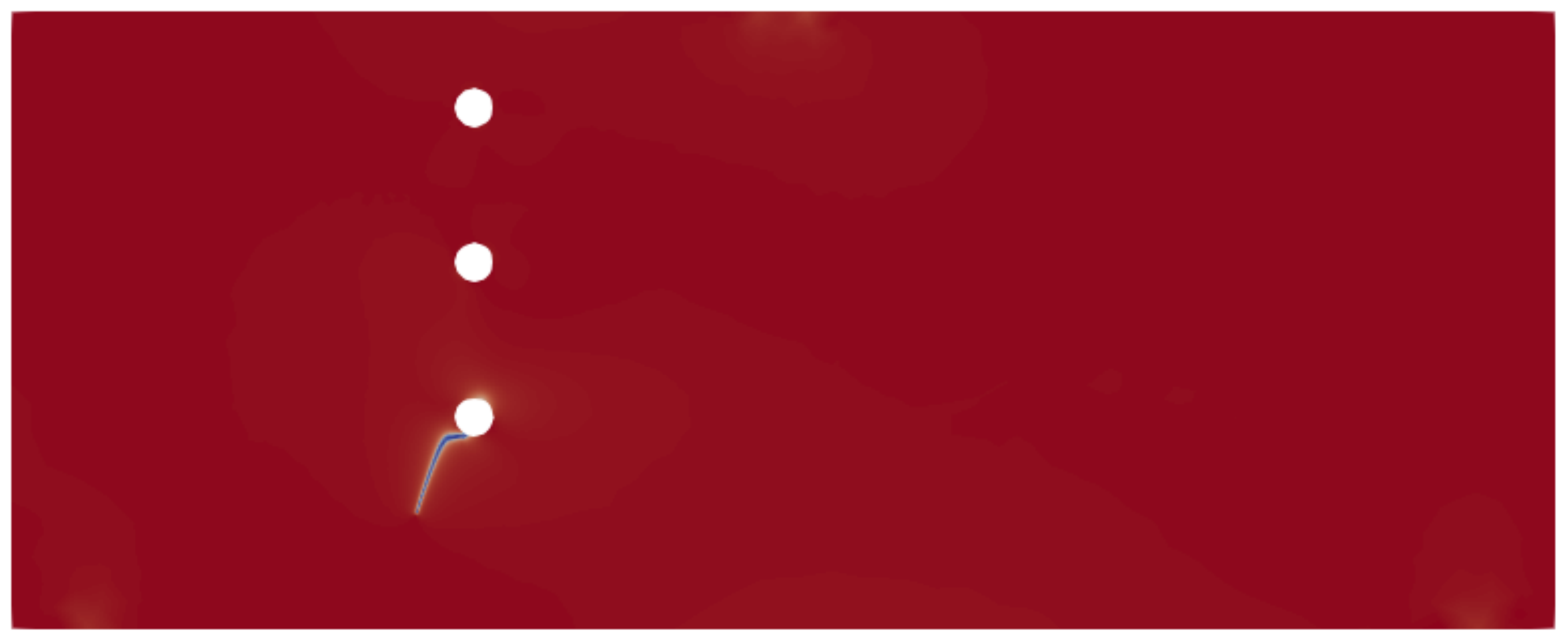}
         \caption{}
     \end{subfigure}
\caption{Damage  profiles for a beam with three holes and a pre-notch defined by $e_1 = 5\, \text{inch}$ and $e_2 = 1.5 \, \text{inch}$ under asymmetric three point bending at different  deformation  stages  shown  in  sub-figures  at  applied  displacement  (a) 0.035 inch (b) 0.0375 inch (c) 0.039 inch and (d) 0.04 inch.}
\label{Fig:AsymTPBtestDamageProfileSpecimenB}
\end{figure}

\begin{figure}[ht!]
     \centering
     \begin{subfigure}[b]{0.33\textwidth}
         \centering
 \includegraphics[width=\textwidth]{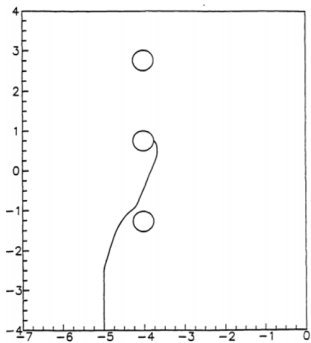}
         \caption{}
     \end{subfigure}
     \begin{subfigure}[b]{0.345\textwidth}
    \centering
 	 \includegraphics[width=\textwidth]{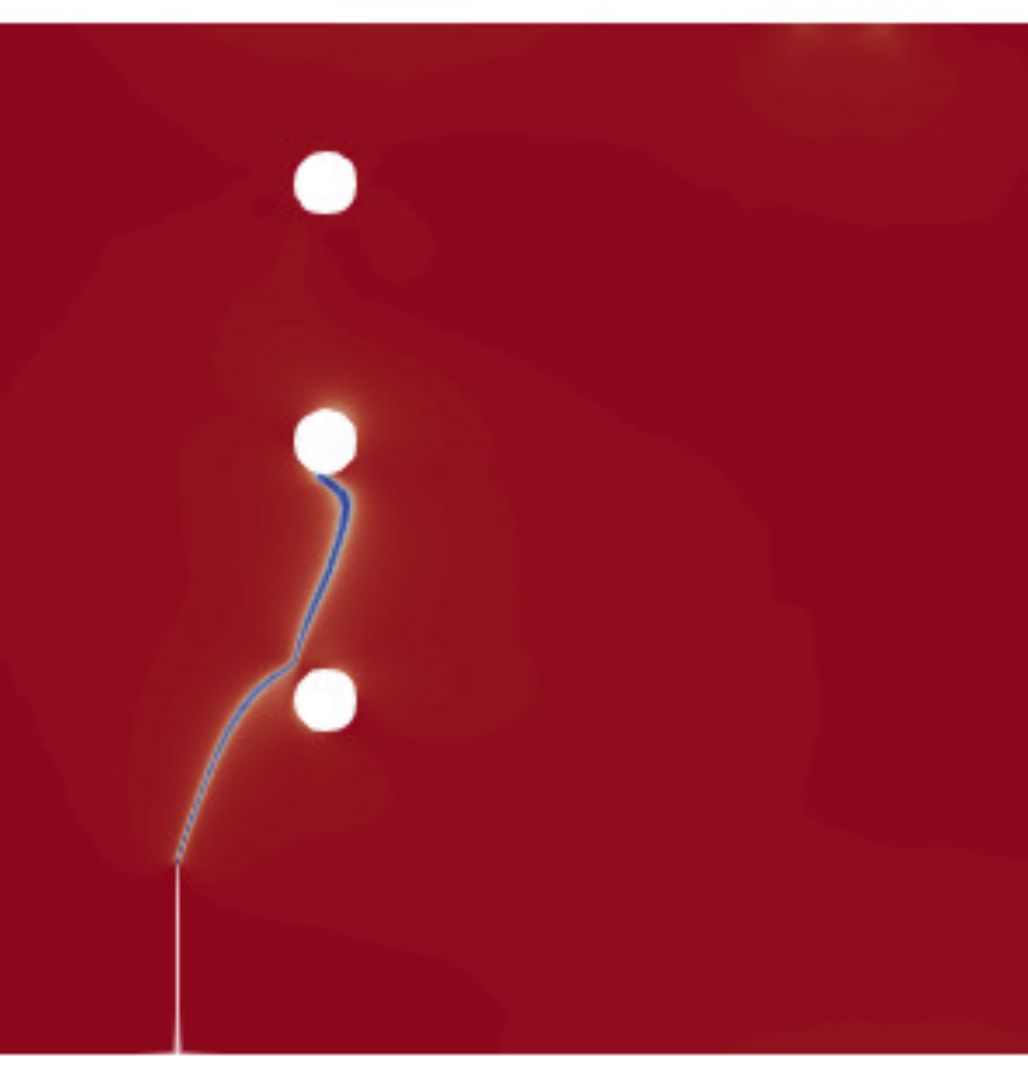}
         \caption{}
     \end{subfigure}
\caption{Comparison of the crack paths for the beam with three holes and a pre-notch defined by $e_1 = 5\, \text{inch}$ and $e_2 = 1.5 \, \text{inch}$ under asymmetric three point bending. Sub-figures (a) and (b) show crack paths for the experimentally observed \cite{ingraffea1990probabilistic} and the numerically predicted, respectively.}
\label{Fig:AsymTPBtestCrackPathSpecimenB}
\end{figure}

\begin{figure}[ht!]
     \centering
 \begin{subfigure}[b]{\textwidth}
         \centering
 \includegraphics[width=\textwidth]{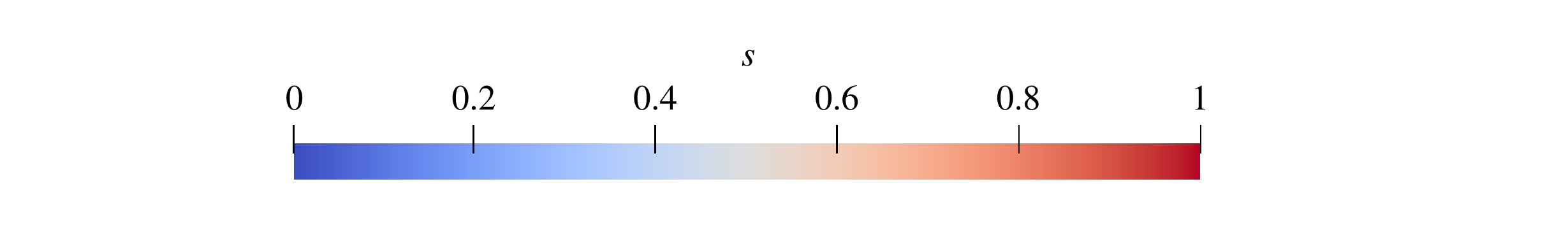}
     \end{subfigure}
     \begin{subfigure}[b]{0.49\textwidth}
         \centering
 \includegraphics[width=\textwidth]{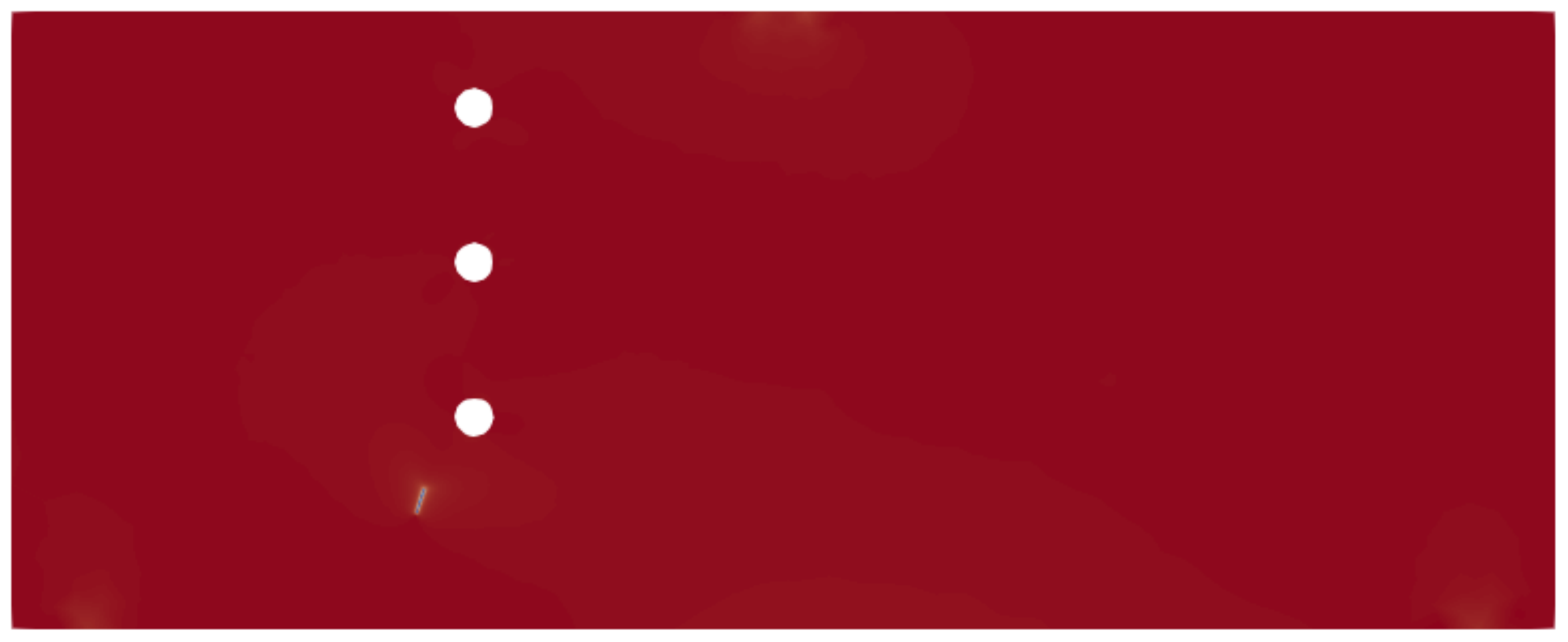}
         \caption{}
     \end{subfigure}
     \begin{subfigure}[b]{0.49\textwidth}
    \centering
 	 \includegraphics[width=\textwidth]{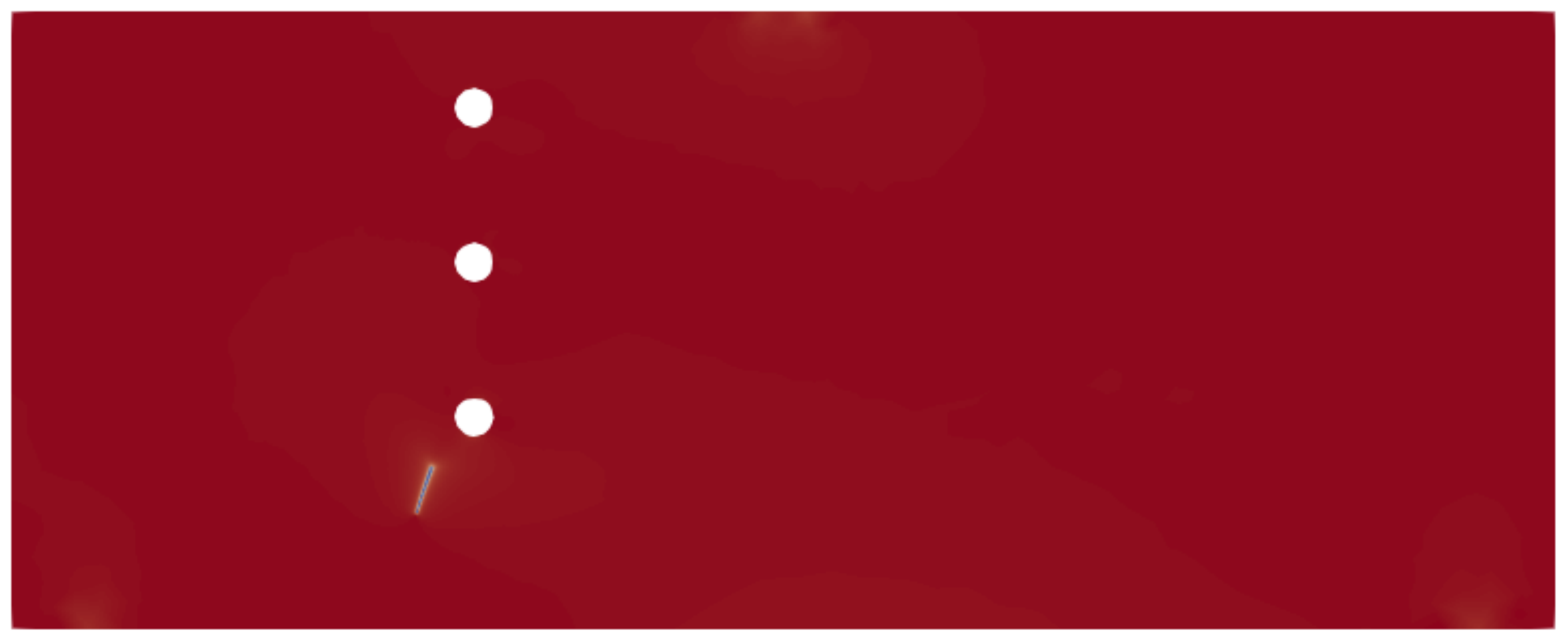}
         \caption{}
     \end{subfigure}
 \begin{subfigure}[b]{0.49\textwidth}
    \centering
 	 \includegraphics[width=\textwidth]{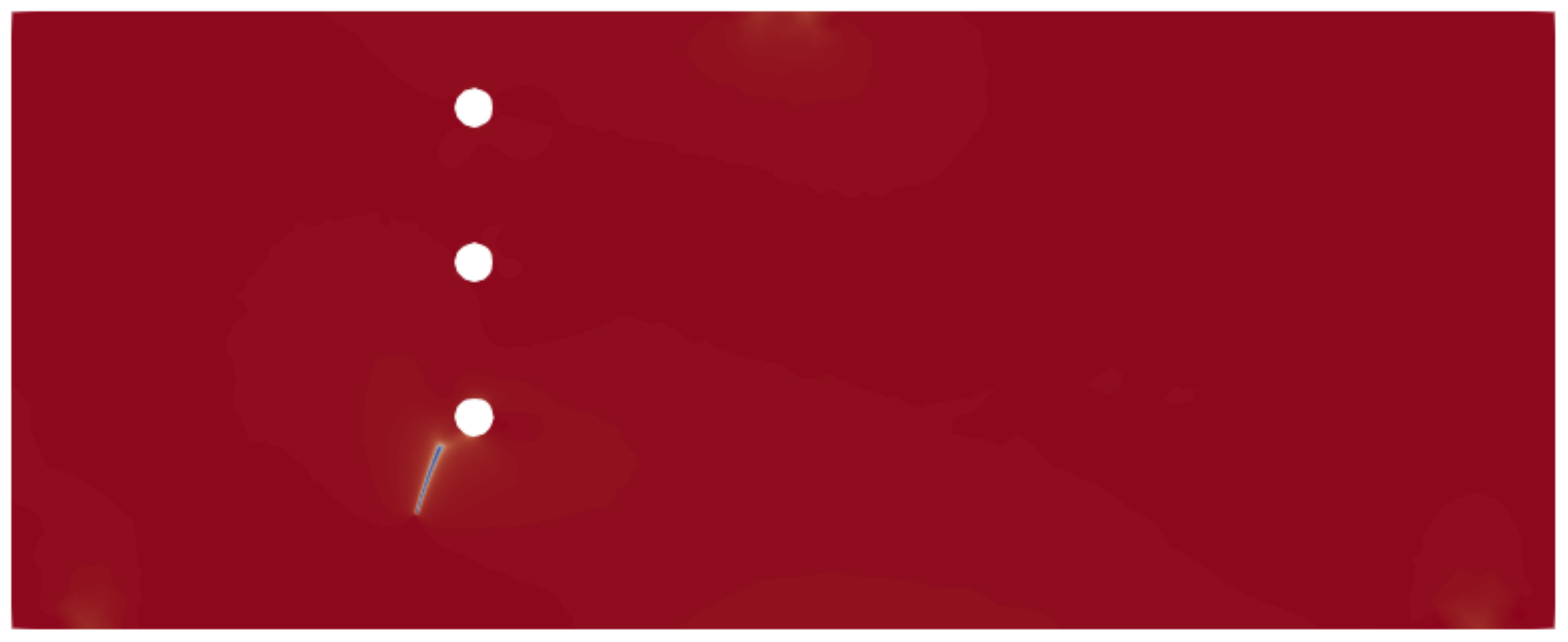}
         \caption{}
     \end{subfigure}
 \begin{subfigure}[b]{0.49\textwidth}
    \centering
 	 \includegraphics[width=\textwidth]{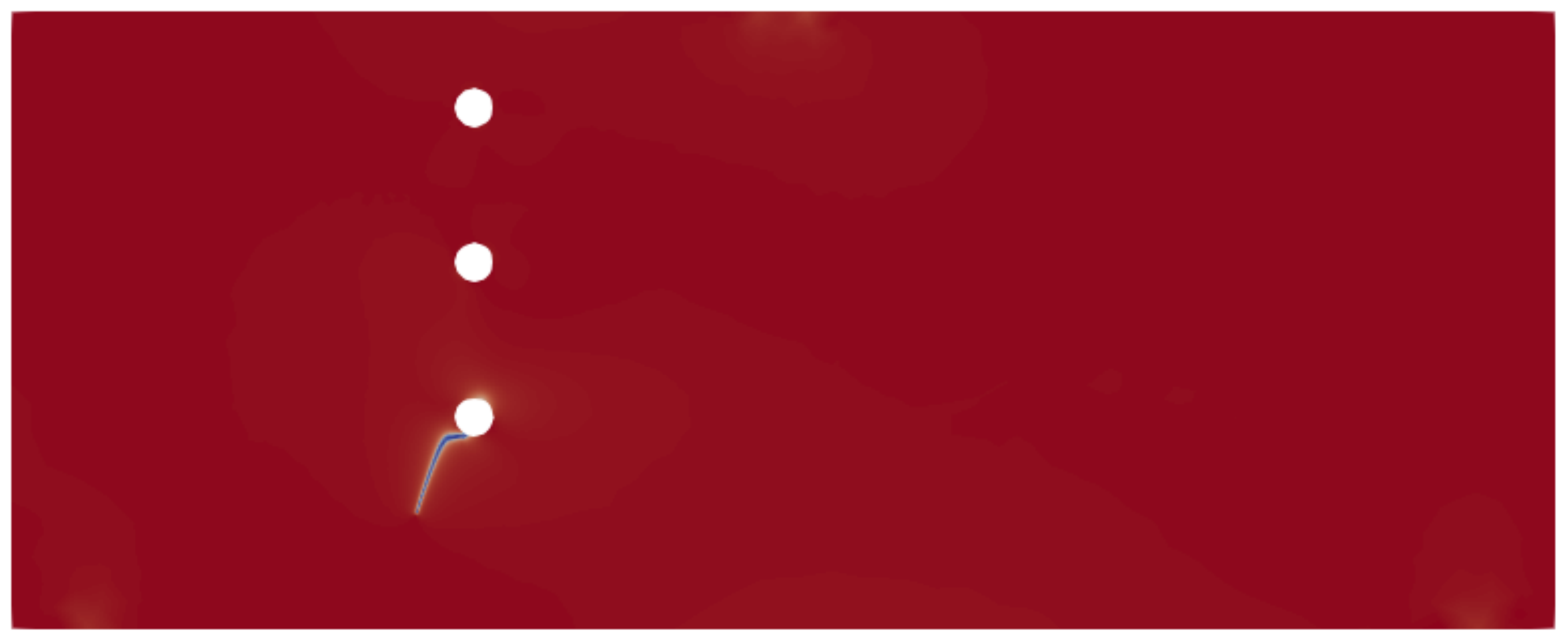}
         \caption{}
     \end{subfigure}
\caption{Damage  profiles for a beam with three holes and a pre-notch defined by $e_1 = 4.75\, \text{inch}$ and $e_2 = 1.5 \, \text{inch}$ under asymmetric three point bending at different  deformation  stages  shown  in  sub-figures  at  applied  displacement  (a) 0.035 inch (b) 0.0375 inch (c) 0.039 inch and (d) 0.04 inch.}
\label{Fig:AsymTPBtestDamageProfileSpecimenC}
\end{figure}

\begin{figure}[ht!]
     \centering
     \begin{subfigure}[b]{0.34\textwidth}
         \centering
 \includegraphics[width=\textwidth]{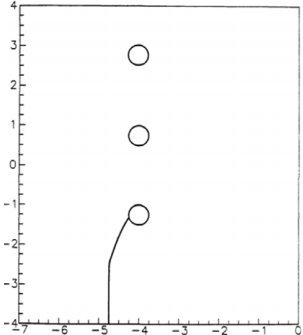}
         \caption{}
     \end{subfigure}
     \begin{subfigure}[b]{0.315\textwidth}
    \centering
 	 \includegraphics[width=\textwidth]{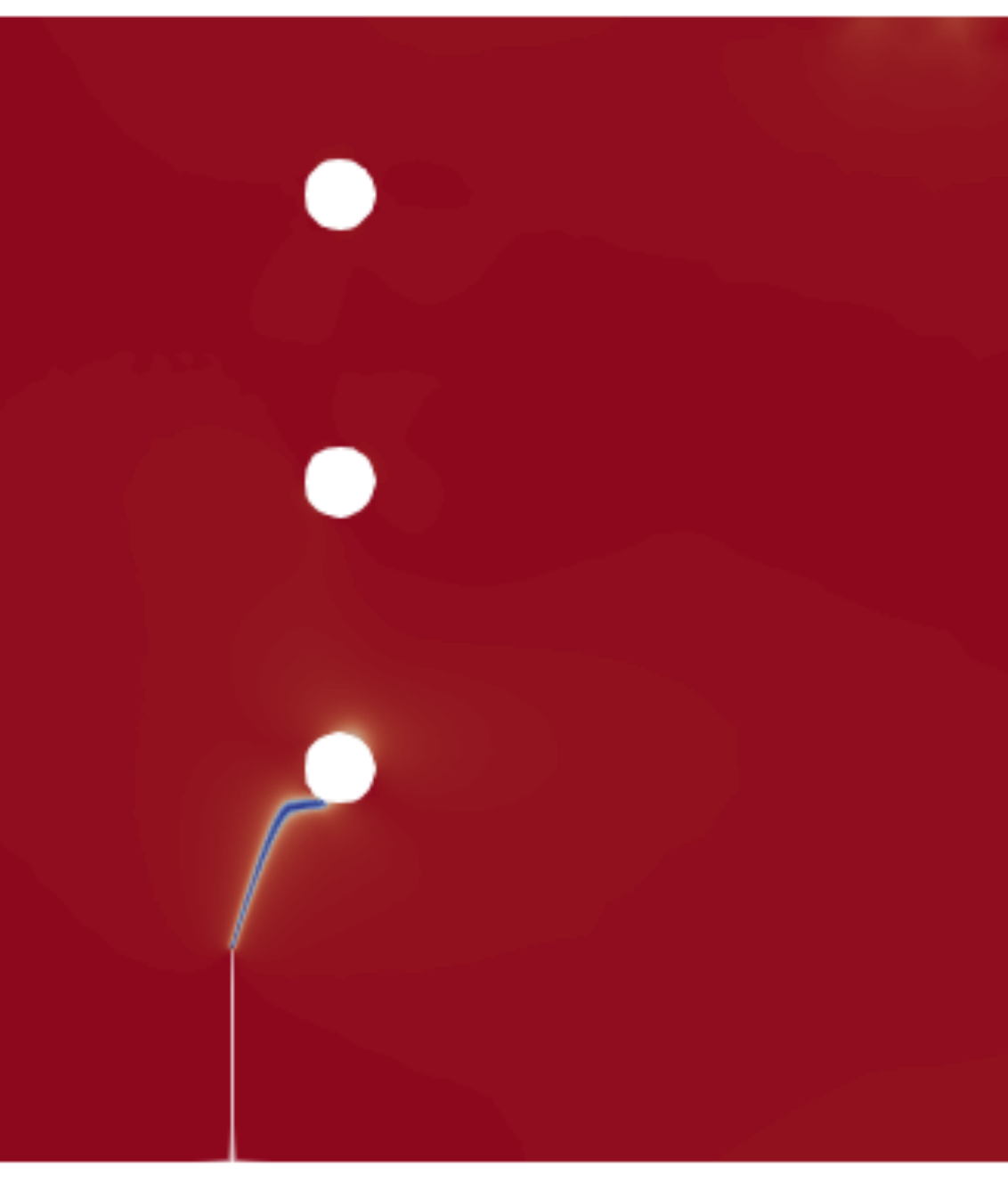}
         \caption{}
     \end{subfigure}
\caption{Comparison of the crack paths for the beam with three holes and a pre-notch defined by $e_1 = 4.75\, \text{inch}$ and $e_2 = 1.5 \, \text{inch}$ under asymmetric three point bending. Sub-figures (a) and (b) show crack paths for the experimentally observed \cite{ingraffea1990probabilistic} and the numerically predicted, respectively.}
\label{Fig:AsymTPBtestCrackPathSpecimenC}
\end{figure}

The developed Julia code for the phase-field model is validated against a set of tests on a notched beam with three holes under asymmetric three-point bending, which were carried out by  Ingraffea and Grigoriu \cite{ingraffea1990probabilistic} and numerically analyzed in Bittencourt {\it{et al.}} \cite{bittencourt1996quasi}. Material parameters are taken as
$E=4.75\times 10^5 \,\text{psi}$, $\nu=0.35$, $G_c = 1.8 \,\text{lbf}/\text{in}$ and $l_s = 0.01 \,\text{inch}$. To verify whether the proposed Julia implementation of phase-field model can predict experimentally observed complex crack patterns, three different conﬁgurations of the specimen characterized by the values $e_1$ and $e _2$ (see Fig. \ref{fig:AsymTPBtestSetUp} for geometry and the boundary conditions, and a finite element mesh used for simulation) are considered.
Prediction of crack path for the beam with three holes and a pre-notch defined by (a) $e_1 = 6\, \text{inch}$ and $e_2 = 1 \, \text{inch}$, (b) $e_1 = 5\,\text{inch}$ and $e_2 = 1.5 \,\text{inch}$ and (c) $e_1 = 4.75\,\text{inch}$ and $e_2 = 1.5\,\text{inch}$ are considered. Displacement control loading (monotonic displacement $\bar{u}_2$ is applied in small increments $\Delta\bar{u}_2$) is considered and the damage profiles for the beam with three holes and a pre-notch defined by (a) $e_1 = 6\, \text{inch}$ and $e_2 = 1 \, \text{inch}$, (b) $e_1 = 5\,\text{inch}$ and $e_2 = 1.5 \,\text{inch}$ and (c) $e_1 = 4.75\,\text{inch}$ and $e_2 = 1.5\,\text{inch}$ at different stages of applied displacement are presented in Fig. \ref{Fig:AsymTPBtestDamageProfileSpecimenA}, Fig. \ref{Fig:AsymTPBtestDamageProfileSpecimenB} and Fig. \ref{Fig:AsymTPBtestDamageProfileSpecimenC}, respectively. As can be seen from Fig. \ref{Fig:AsymTPBtestCrackPathSpecimenA}, Fig. \ref{Fig:AsymTPBtestCrackPathSpecimenB} and Fig. \ref{Fig:AsymTPBtestCrackPathSpecimenC}, the proposed Julia implementation shows a very good prediction of the experimentally observed crack paths which are very sensitive to the height and relative location of the pre-notch. Remarkably, the proposed implementation reproduces the intricate deviation of crack path due to the local stress concentration around the bottom hole which is experimentally observed (see Fig. \ref{Fig:AsymTPBtestCrackPathSpecimenB}).  

\clearpage

\section{Concluding Remarks}
\label{sec:conclusion}

The present study has provided a novel numerical implementation of a thermodynamically consistent phase-field model for brittle fracture using an open-source ﬁnite element toolbox, Gridap in Julia. The proposed implementation is validated against a few numerical and experimental results available in the literature. As the proposed implementation is available with an open-source
license, it may eliminate the technical barrier for practitioners and researchers who are interested to explore the phase-field model for solving a wide range of brittle fracture problems. Moreover, the proposed implementation will expose the users to many built-in packages of Julia that may be useful for researchers who want to extend the proposed implementation for the case of ductile fracture or other applications. Most
importantly, the availability of an open-source code that is compact, user friendly, highly efficient, and accessible to everyone will allow a third-party verification and essentially establish a high standard for efficient open-source code development.

\section*{Data accessibility} The present work does not generate any experimental data. Julia scripts used as the source codes for symmetric three-point bending test are provided in Section \ref{sec:JuliaImplementation} of this article. Jupyter notebook files for the proposed Julia implementation of a phase-field model can be downloaded from \href{https://github.com/MdMasiurRahaman/PhaseFieldModelForBrittleFracture/blob/main/BeamWithNotchThreePointBendingFinal.ipynb}{Julia code for phase-field model}.

\section*{Declaration of Competing Interest}
The author of this article declares that he has no conflict of interest.

\section*{Acknowledgments}
The author of this article gratefully acknowledge support from the Indian Institute of Technology Bhubaneswar under the grant SP107.

\clearpage

\appendix

\section{Julia code to create finite element mesh file for a notched beam under symmetric three point bending}
\label{Sec:AppendixFEmeshSymTPBtest}

One can generate finite element mesh file in Julia by using the GMSH mesh generator, which can be loaded in Julia by writing the following line.
\begin{lstlisting}[language=C,backgroundcolor=\color{gray!5!white},showstringspaces=false,breaklines=true,basicstyle=\ttfamily]
using Gmsh: gmsh
\end{lstlisting}
One can create the mesh file ``BeamWithNotchSymThreePtBending.msh" using the following lines in Julia.
\begin{lstlisting}
const L = 8.0
const LL = 0.475.*L
const LR = 0.525.*L

const H = 2.0
const CH = 0.4 #Crack height
const CW = 0.2 #Crack Width
const ls = 0.03
const hfc = ls/100 #Mesh size parameter
const hf = ls/2.1 #Mesh size parameter

const h = 100*hf #Mesh size parameter

gmsh.initialize()
gmsh.option.setNumber("General.Terminal", 1)
gmsh.model.geo.addPoint((L/2)+(CW/2), 0.0 , 0.0, hf,1)
gmsh.model.geo.addPoint(L, 0.0, 0.0, h, 2)
gmsh.model.geo.addPoint(L, H, 0.0, h, 3)
gmsh.model.geo.addPoint(LR, H, 0.0, hf, 4)
gmsh.model.geo.addPoint(LL, H, 0.0, hf, 5)
gmsh.model.geo.addPoint(0.0, H, 0.0, h, 6)
gmsh.model.geo.addPoint(0.0, 0.0, 0.0, h, 7)
gmsh.model.geo.addPoint((L/2)-(CW/2), 0.0 , 0.0, hf,8)
gmsh.model.geo.addPoint((L/2), CH , 0.0, hfc, 9)   

gmsh.model.geo.addLine(1, 2, 1)
gmsh.model.geo.addLine(2, 3, 2)
gmsh.model.geo.addLine(3, 4, 3)
gmsh.model.geo.addLine(4, 5, 4)
gmsh.model.geo.addLine(5, 6, 5)
gmsh.model.geo.addLine(6, 7, 6)
gmsh.model.geo.addLine(7, 8, 7)
gmsh.model.geo.addLine(8, 9, 8)
gmsh.model.geo.addLine(9, 1, 9) 
gmsh.model.geo.addCurveLoop([1,2,3,4,5,6,7,8,9],1) 

gmsh.model.geo.addPlaneSurface([1], 1)

gmsh.model.addPhysicalGroup(2, [1],1)

gmsh.model.addPhysicalGroup(1, [4],1)
gmsh.model.addPhysicalGroup(0, [7],2)
gmsh.model.addPhysicalGroup(0, [2],3)

gmsh.model.setPhysicalName(2, 1, "Domain")


gmsh.model.setPhysicalName(1, 1, "LoadLine")
gmsh.model.setPhysicalName(0, 2, "LeftSupport")
gmsh.model.setPhysicalName(0, 3, "RightSupport")

gmsh.model.mesh.field.add("Box", 10)
gmsh.model.mesh.field.setNumber(10, "VIn", hf)
gmsh.model.mesh.field.setNumber(10, "VOut", h)
gmsh.model.mesh.field.setNumber(10, "XMin", (L/2)-CW)
gmsh.model.mesh.field.setNumber(10, "XMax", (L/2)+CW)
gmsh.model.mesh.field.setNumber(10, "YMin", 0)
gmsh.model.mesh.field.setNumber(10, "YMax", H)
gmsh.model.mesh.field.setAsBackgroundMesh(10)

gmsh.model.geo.synchronize()
gmsh.model.mesh.generate(2)
gmsh.write("BeamWithNotchSymThreePtBending.msh")
gmsh.finalize()
\end{lstlisting}

\section{Julia code to create finite element mesh file for a notched beam with three holes under asymmetric three point bending}
\label{Sec:AppendixFEmeshAsymTPBtest}
One can create the mesh file ``AsymThreePtBending.msh" by using the following lines in Julia.
\begin{lstlisting}[language=C,backgroundcolor=\color{gray!5!white},showstringspaces=false,breaklines=true,basicstyle=\ttfamily]
using Gmsh: gmsh

const L = 20.0
const LL = 0.475.*L
const LR = 0.525.*L

const H = 8.0
const CH = 1.5 #Crack height
const CW = L/2000 #Crack Width
const $e_1$ = 5.15
const CP = L/2 - $e_1$
const SD = 1.0
const HP = 6.0
const HR = 0.25
const HH1 = 2.75
const HH2 = 4.75
const HH3 = 6.75
const ls = 0.01

const hfc = ls/50  #Mesh size parameter
const hf = ls/2.1 #Mesh size parameter
const hfl = 50*hf  #Mesh size parameter
const hfh = hf  #Mesh size parameter

const h = 100*hf #Mesh size parameter

$\theta$ = $\pi$/180
cr1 = CP+(CW/2) + HP*tan($\theta$)
cr2 = CP-(CW/2) + HP*tan($\theta$)

const FMR = 40*ls 

gmsh.initialize()
gmsh.option.setNumber("General.Terminal", 1)

p1 = gmsh.model.geo.addPoint(CP+(CW/2), 0.0 , 0.0, h)
p2 = gmsh.model.geo.addPoint(L-SD, 0.0, 0.0, h)
p3 = gmsh.model.geo.addPoint(L, 0.0, 0.0, h)
p4 = gmsh.model.geo.addPoint(L, H, 0.0, h)
p5 = gmsh.model.geo.addPoint(LR, H, 0.0, hfl)
p6 = gmsh.model.geo.addPoint(LL, H, 0.0, hfl)
p7 = gmsh.model.geo.addPoint(0.0, H, 0.0, h)
p8 = gmsh.model.geo.addPoint(0.0, 0.0, 0.0, h)
p9 = gmsh.model.geo.addPoint(SD, 0.0, 0.0, h)
p10 = gmsh.model.geo.addPoint(CP-(CW/2), 0.0, 0.0, h)
p11 = gmsh.model.geo.addPoint(CP-(CW/2), CH, 0.0, hfc) 
p12 = gmsh.model.geo.addPoint(CP+(CW/2), CH, 0.0, hfc)  

l1 = gmsh.model.geo.addLine(p1, p2)
l2 = gmsh.model.geo.addLine(p2, p3)
l3 = gmsh.model.geo.addLine(p3, p4)
l4 = gmsh.model.geo.addLine(p4, p5)
l5 = gmsh.model.geo.addLine(p5, p6)
l6 = gmsh.model.geo.addLine(p6, p7)
l7 = gmsh.model.geo.addLine(p7, p8)
l8 = gmsh.model.geo.addLine(p8, p9)
l9 = gmsh.model.geo.addLine(p9, p10) 
l10 = gmsh.model.geo.addLine(p10, p11) 
l11 = gmsh.model.geo.addLine(p11, p12) 
l12 = gmsh.model.geo.addLine(p12, p1) 
cl1 = gmsh.model.geo.addCurveLoop([l1,l2,l3,l4,l5,l6,l7,l8,l9,l10,l11,l12]) 


p13 = gmsh.model.geo.addPoint(HP-HR, HH1, 0.0, hfh)
p14 = gmsh.model.geo.addPoint(HP+HR, HH1, 0.0, hfh)
p15 = gmsh.model.geo.addPoint(HP, HH1, 0.0, hfh)

ca1 = gmsh.model.geo.addCircleArc(p13, p15, p14)
ca2 = gmsh.model.geo.addCircleArc(p14, p15, p13)
cl2 = gmsh.model.geo.addCurveLoop([ca1,ca2])

p16 = gmsh.model.geo.addPoint(HP-HR, HH2, 0.0, hfh)
p17 = gmsh.model.geo.addPoint(HP+HR, HH2, 0.0, hfh)
p18 = gmsh.model.geo.addPoint(HP, HH2, 0.0, hfh)

ca3 = gmsh.model.geo.addCircleArc(p16, p18, p17)
ca4 = gmsh.model.geo.addCircleArc(p17, p18, p16)
cl3 = gmsh.model.geo.addCurveLoop([ca3,ca4])

p19 = gmsh.model.geo.addPoint(HP-HR, HH3, 0.0, hfh)
p20 = gmsh.model.geo.addPoint(HP+HR, HH3, 0.0, hfh)
p21 = gmsh.model.geo.addPoint(HP, HH3, 0.0, hfh)

ca5 = gmsh.model.geo.addCircleArc(p19, p21, p20)
ca6 = gmsh.model.geo.addCircleArc(p20, p21, p19)
cl4 = gmsh.model.geo.addCurveLoop([ca5,ca6])

ps1 = gmsh.model.geo.addPlaneSurface([cl1,-cl2,-cl3,-cl4])

pg1 = gmsh.model.addPhysicalGroup(2, [ps1])

pg2 = gmsh.model.addPhysicalGroup(1, [l5])
pg3 = gmsh.model.addPhysicalGroup(0, [p9])
pg4 = gmsh.model.addPhysicalGroup(0, [p2])

gmsh.model.setPhysicalName(2, pg1, "Domain")


gmsh.model.setPhysicalName(1, pg2, "LoadLine")
gmsh.model.setPhysicalName(0, pg3, "LeftSupport")
gmsh.model.setPhysicalName(0, pg4, "RightSupport")

p22 = gmsh.model.geo.addPoint(CP-(CW/2), 0.8*CH, 0.0, hf) 
p23 = gmsh.model.geo.addPoint(CP+(CW/2), 0.8*CH, 0.0, hf) 

p24 = gmsh.model.geo.addPoint(HP, cr1, 0.0, hf)
p25 = gmsh.model.geo.addPoint(HP, cr2, 0.0, hf)

l13 = gmsh.model.geo.addLine(p22, p24)
l14 = gmsh.model.geo.addLine(p23, p25)


gmsh.model.mesh.field.add("Distance", 1)
gmsh.model.mesh.field.setNumbers(1, "EdgesList", [l13,l14])

gmsh.model.mesh.field.add("Threshold", 2)
gmsh.model.mesh.field.setNumber(2, "IField", 1)
gmsh.model.mesh.field.setNumber(2, "LcMin", hf)
gmsh.model.mesh.field.setNumber(2, "LcMax", h)
gmsh.model.mesh.field.setNumber(2, "DistMin", FMR)
gmsh.model.mesh.field.setNumber(2, "DistMax", 1.5*FMR)

l15 = gmsh.model.geo.addLine(p24,p6)
l16 = gmsh.model.geo.addLine(p23, p5)

gmsh.model.mesh.field.add("Distance", 3)
gmsh.model.mesh.field.setNumbers(3, "EdgesList", [l15,l16])

gmsh.model.mesh.field.add("Threshold", 4)
gmsh.model.mesh.field.setNumber(4, "IField", 3)
gmsh.model.mesh.field.setNumber(4, "LcMin", hfl)
gmsh.model.mesh.field.setNumber(4, "LcMax", h)
gmsh.model.mesh.field.setNumber(4, "DistMin", FMR)
gmsh.model.mesh.field.setNumber(4, "DistMax", 1.5*FMR)


gmsh.model.mesh.field.add("Min",5)
gmsh.model.mesh.field.setNumbers(5, "FieldsList",[2,4])
gmsh.model.mesh.field.setAsBackgroundMesh(5)
gmsh.model.geo.synchronize()
gmsh.model.mesh.generate(2)
gmsh.write("AsymThreePtBending.msh")
gmsh.finalize()
\end{lstlisting}






\clearpage
\bibliographystyle{elsarticle-num-names}
\bibliography{References.bib}







\end{document}